\documentclass[10pt,twocolumn]{aastex61}
\received{xxx}
\revised{xxx}
\accepted{xxx}

\usepackage{lineno}

\submitjournal{ApJ}
\shorttitle{Gravitational Waves from Accreting Neutron Stars in Common Envelopes}
\shortauthors{Holgado et al.}
\usepackage{mathrsfs,amsmath,float}
\usepackage{verbatim,enumerate,soul,stix}
\graphicspath{{./figures/}}
\begin{document}
\title{Gravitational Waves from Accreting Neutron Stars \\ undergoing Common-Envelope Inspiral}
\correspondingauthor{A.~Miguel~Holgado}
\email{holgado2@illinois.edu}
\author{A.~Miguel~Holgado}
\affiliation{Department of Astronomy, University of Illinois at Urbana-Champaign, Urbana, IL 61801, USA}
\affiliation{National Center for Supercomputing Applications, University of Illinois at Urbana-Champaign, Urbana, IL 61801, USA}
%
\author{Paul~M.~Ricker}
\affiliation{Department of Astronomy, University of Illinois at Urbana-Champaign, Urbana, IL 61801, USA}
\affiliation{National Center for Supercomputing Applications, University of Illinois at Urbana-Champaign, Urbana, IL 61801, USA}
\author{E.~A. Huerta}
\affiliation{National Center for Supercomputing Applications, University of Illinois at Urbana-Champaign, Urbana, IL 61801, USA}
\begin{abstract}

The common envelope phase is a likely formation channel for close binary systems containing compact objects. Neutron stars in common envelopes accrete at a fraction of the Bondi-Hoyle-Lyttleton accretion rate, since the stellar envelope is inhomogeneous, but may still be able to accrete at hyper-critical rates (though not enough to become black holes).
We show that common envelope systems consisting of a neutron star with a massive primary may be gravitational wave sources detectable in the Advanced LIGO band as far away as the Magellanic Clouds.
To characterize their evolution, we perform orbital integrations using 1D models of $12 M_\odot$ and $20 M_\odot$ primaries, considering the effects of density gradient on the accretion onto the NS and spin evolution.
From the range of possible accretion rates relevant to common envelope evolution, we find that these systems may be significantly louder gravitational wave sources than low-mass X-ray binaries like Sco X-1, which are currently the target of directed searches for continuous GWs.
We also find that their strain amplitude signal may allow for novel constraints on the orbital separation and inspiral timescale in common envelopes when combined with pre-common envelope electromagnetic observations.
\end{abstract}
\section{Introduction} \label{sec:intro}
Neutron stars (NSs) and black holes (BHs) are the end points of massive stellar evolution. 
With the recent detection of binary black holes (BBHs) through gravitational wave (GW) emission \citep{DI:2016,secondBBH:2016,bbhswithligo:2016,ligo_scientific_and_virgo_collaboration_gw170104:_2017,ligo_scientific_collaboration_and_virgo_collaboration_gw170814:_2017}, we have a powerful new probe of BH physics, and we expect that compact binaries with NS components will also be detected in the very near future \citep{scenarioligo:2016LRR}. 
A worldwide network of kilometer-scale GW detectors formed by the Advanced Laser-Interferometer Gravitational-Wave Observatory (aLIGO) detectors~\citep{LSC:2015}, Advanced Virgo~\citep{Virgo:2015}, Kagra~\citep{Hiroshe:2014}, and LIGO-India~\citep{Unni:2013} working at design sensitivity may enable routine detection of GW sources. 
The resulting catalogs will constrain the numbers and masses of ultra-compact astrophysical objects and will provide unique insights into the stellar evolution processes that lead to their formation \citep{Pod:1992ApJ,Heger:2003ApJ,bogomazov_evolution_2007,Langer:2012ARA,Post:2014LRR}.
\par
The detection of BBHs with aLIGO confirmed population synthesis models that had predicted that BHs would be the first GW sources to be detected with ground-based GW detectors, and in particular that massive stellar-mass BHs should be found \citep{bel:2010ApJ,belczynski_first_2016,stevenson_formation_2017,belczynski_gw170104_2017}. 
It also led to the development of numerous models proposing a wide variety of formation channels \citep[e.g.,][]{Anto:2015arXiv,D9:2016,scenarioligo:2016LRR,Carl:2016arXiv,marchant:2016,deMink:2016MNRAS,racca:2016PhRvD,mandic:2016PhRvL}. 
These formation scenarios, varying from plausible to speculative, produce stellar mass BHs in a sufficiently close orbit that GW emission can drive the binary to merger within a Hubble time.
The channel originally proposed by Belczynski \textit{et al.}\ uses common envelope (CE) evolution to achieve the needed orbital separation \citep{belczynski_first_2016,Kruckow:2016AA}. 
This phase of binary stellar evolution involves a primary star enveloping its companion, so that the two stellar cores share a single envelope.
Gravitational drag between the cores and the common envelope causes the binary to inspiral, thus reducing the orbital separation.
As observations suggest that more than 70\% of massive stars exchange matter with a nearby companion \citep{Sana:2012}, channels involving CE evolution are likely to be an important contributor to BBH populations and other compact binary systems \citep[e.g.,][]{taam_common_2000}.  
\par
The CE process, as first introduced by \cite{paczynski_common_1976}, 
 is an outstanding problem in binary stellar evolution with many open questions.
Significant progress has been made toward a more complete 3D description of the orbital dynamics 
\citep{terman_double-core_1994,terman_double_1995,terman_double-core_1996,rasio_formation_1996,ivanova_common_2011,ricker_amr_2012,passy_simulating_2012,nandez_v1309_2014,nandez_testing_2015,ivanova_role_2015,ohlmann_magnetic_2016,ohlmann_hydrodynamic_2016,staff_hydrodynamic_2016,iaconi_effect_2017}.
However, a lack of physical data on CE systems has made it difficult to constrain models of CE evolution\footnote{ 
See \cite{ivanova_common_2013} for a thorough review of the CE problem, including a summary of recent advances in the field and prospects for future developments in CE theory.}.
Additional sources of energy appear to be required to prevent merger within the envelope, with recombination being one of several putative candidates \citep[e.g.,][]{nandez_testing_2015}.
\par
The CE phase may play a role in forming systems such as binary white dwarfs (BWDs), binary neutron stars (BNSs), black hole-neutron star  (NSBH) binaries, and BBHs~\citep{Kruckow:2016AA}.
The scenario we consider in this work is most relevant to BNSs and NSBHs with high-mass X-ray binary progenitors that undergo CE evolution, wherein the NS survives the CE phase.
We take survival to mean that the NS both ejects the envelope, preventing the formation of a Thorne-\.{Z}ytkow object, and does not accrete enough mass to collapse to a BH.
During the CE phase, accretion onto a NS companion may be hyper-critical \citep{houck_steady_1991,armitage_black_2000}.
However, the accretion rate onto compact companions during CE evolution is greatly suppressed from the Bondi-Hoyle-Littleton (BHL) expectation \citep{ricker_interaction_2008}.
\cite{macleod_asymmetric_2015}, \cite{macleod_accretion-fed_2015}, and \cite{macleod_common_2017} found, using 3D wind-tunnel simulations of accretion from flows including a transverse density gradient, that the degree of suppression increases with the magnitude of the density gradient. 
Considering these results with density gradients typical of stellar envelopes, they concluded that NSs should survive the CE phase. 
\cite{macleod_accretion-fed_2015}, hereafter MR15, used a fit to these results to model the CE evolution of a $1.33M_\odot$ NS with $12 - 20 M_\odot$ supergiant companions and found the amount of accretion to be well below the threshold needed for collapse to a BH.
\par
A system with a finite, changing quadrupole moment produces gravitational radiation.
Such systems include accreting NSs, which can be spun up to millisecond periods and can exhibit asymmetries that introduce a changing quadrupole moment \citep[e.g.,][]{wagoner_gravitational_1984,lai_gravitational_1995,bildsten_gravitational_1998,melatos_gravitational_2007,horowitz_gravitational_2010,patruno_gravitational_2012,lasky_gravitational_2015}.
\cite{piro_gravitational_2012} (hereafter PT12) presented an analytic model for the GW emission from an accreting NS for a supernova fallback scenario.
In the PT12 model, the NS is treated as a Maclaurin spheroid, and the strain amplitude is estimated assuming that GW torque efficiently counterbalances accretion torque once the NS becomes secularly unstable. 
Observed NS spin frequencies in X-ray binaries lie below the secular instability threshold \citep{papitto_spin_2014,patruno_spin_2017}. 
However, quadrupole moments generated by other mechanisms, such as the temperature dependence of electron capture reactions \citep{bildsten_gravitational_1998,ushomirsky_deformations_2000} or burial of the magnetic field by accreted material \citep{payne_burial_2004,melatos_gravitational_2005}, can easily exceed the levels needed to produce torque balance at the observed spins.
In the context of Thorne-\.{Z}ytkow object formation, \cite{nazin_gravitational-wave_1997} considered GW emission within the vicinity of the inspiralling NS generated by anisotropic neutrino emission in the presence of strong magnetic fields, i.e., $B \gtrsim 10^{12} \, {\rm G}$, where the frequency is modulated by the NS spin frequency.
The strain amplitudes they estimate are $\sim3{\times}10^{-28}$ at a distance of $10 \, {\rm kpc}$.
\par 
In this paper, we use the MR15 model for CE evolution to estimate the GW emission from crustal deformations of accreting NSs undergoing CE inspiral and the resulting spin-up.
We include a prescription to account for the effects of NS crust melting during inspiral, which decreases the strain amplitude as the NS accretes at hyper-critical rates.  
We propose that a direct detection of GWs from a NS-CE binary may constrain estimates of the orbital separation and decay rate of the binary given a model for the primary star, helping to address key questions in common envelope evolution such as the condition for merger versus stabilization as a close binary, the mechanisms responsible for envelope ejection, and the efficiency of envelope ejection.
\par
We use Scorpius X-1 (Sco X-1), a low-mass X-ray binary (LMXB) system, as a reference continuous GW source.
The compact companion in this system is an accreting NS that has been observed to have an X-ray luminosity $L_{\rm X} \approx 2.3 {\times} 10^{38} \ {\rm erg} \  {\rm s}^{-1}$, which is consistent with the NS accreting at its Eddington limit \citep[e.g.,][]{bradshaw_high-resolution_1999}.
At a distance  $D \approx 2.8~{\rm kpc}$, Sco X-1 is thought to be the loudest continuous GW source in the aLIGO band, which covers the frequency range  $10 \  {\rm Hz} \lesssim f \lesssim 1 \  {\rm kHz}$  \citep[e.g.,][]{DI:2016,secondBBH:2016,bbhswithligo:2016}.
\par
To estimate the strain-amplitude evolution of a NS undergoing CE inspiral, we construct models for the CE structure, NS orbital evolution, NS accretion rate, and GW emission from the spinning NS.
These models are simplified and primarily serve to demonstrate how the strain amplitude and NS spin may evolve with the accretion rate as the NS inspirals.
Performing global 3D simulations will be the subject of future work and will provide more accurate characterizations of the orbital evolution and the resulting strain amplitude. 
Our estimates for the accretion rates and strain amplitudes are upper limits, given the simplifying assumptions we use.
\par
It is possible for NSs to accrete above the Eddington limit, i.e., to accrete in a hyper-critical regime.
The lower limit on the accretion rate of hyper-accreting NSs is set by thermal neutrino cooling.
Using the latest upper limits for the maximum quadrupole moment that the NS crust can sustain, we estimate the greatest distance that hyper-accreting NSs would be detected at strain amplitudes comparable to Sco X-1.
\par
In \S~\ref{sec:ce}, we provide a brief review of aspects of CE theory needed for this work.
In \S~\ref{sec:acc}, we discuss NS accretion and the physical regimes relevant to X-ray binaries and common envelopes.
In \S~\ref{sec:gw}, we review GW emission from accreting NSs and discuss the torque balance assumption.
In \S~\ref{sec:methods}, we describe the detailed model we use to predict the GW signal from NS-CE systems.
In \S~\ref{sec:res}, we present the orbital evolution of our models.
We discuss in \S~\ref{sec:disc} how orbital separation and decay rate can be constrained and consider the detectability of NS-CE events with aLIGO.
We conclude in \S~\ref{sec:con} with a summary of our results.
\section{Common Envelope Preliminaries} \label{sec:ce}
\subsection{Evolutionary sequence}
A star in a close binary can expand over the course of its stellar evolution to the point where it envelopes its companion star, thus initiating a CE phase.
It is thought that the CE phase involves a contact phase, followed by a plunge-in phase in which the companion undergoes a fast inspiral, and finally a self-regulated phase \citep{ivanova_common_2013}.
If enough orbital energy is dissipated in the envelope, the system transitions to a post-CE phase in which the remainder of the envelope is ejected. 
In this paper, we neglect the initial plunge, which typically lasts 1-2 orbits in 3D simulations, and treat the entire orbital evolution as self-regulated and driven by local gravitational drag in the vicinity of the NS companion. The adequacy of this approximation remains to be shown for the binary system parameters we consider here. We are undertaking this task using 3D simulations that will be presented in a future paper.
%
\subsection{The $\alpha_{\rm CE}$ parameter}
The energy formalism is commonly used to parameterize CE evolution \citep[e.g.,][]{webbink_double_1984} using the efficiency $\alpha_{\rm CE}$ with which orbital energy release unbinds the envelope.
The binding energy of the envelope, $E_{\rm bind}$, is then related to the orbital energy dissipated over the course of the inspiral, $\Delta E_{\rm orb}$, via
\begin{equation}
E_{\rm bind} = \alpha_{\rm CE} \Delta E_{\rm orb} = 
\alpha_{\rm CE}\left(\frac{G M_{\rm core} M_{\rm NS}}{2 a_{\rm f}} - \frac{G M_1 M_{\rm NS}}{2 a_{\rm i}}\right) \ ,
\end{equation} 
where $M_1$ is the mass of the primary, $M_{\rm core}$ is the mass of the primary's core, $M_{\rm NS}$ is the mass of the NS companion, $a_{\rm i}$ is the initial orbital separation (taken here to be the radius of the primary), and $a_{\rm f}$ is the final separation.
The core-envelope split for the primary is an uncertainty, such that $M_{\rm core}$ usually includes both the mass of the primary's core and the leftover envelope that remains bound to it. 
Some observed systems require $\alpha_{\rm CE} > 1$, implying that additional sources of energy contribute to unbinding the envelope\footnote{See \cite{marco__2011} for a detailed review of the energy formalism.}.
\section{Accreting Neutron Stars in Binary Systems} \label{sec:acc}
\subsection{Eddington-limited accretion in LMXBs}
Neutron stars in low-mass X-ray binaries can accrete from their companion stars at rates as high as the Eddington accretion rate, given by
\begin{equation}
{\dot M}_{\rm Edd}  = \frac{4 \pi G M_{\rm NS}}{\zeta c\kappa_{\rm T}} \ ,
\end{equation}
where $\kappa_{\rm T}$ is the Thomson scattering opacity and $\zeta$ is the efficiency with which the mass accretion rate powers the luminosity, typically taken to be $\zeta \sim 0.1$. 
We re-express the Eddington accretion limit in terms of typical NS parameters as
\begin{equation}
{\dot M}_{\rm Edd}  
\approx 3.5{\times}10^{-8} \left(\frac{M_{\rm NS}}{1.33 M_\odot}\right)  \left(\frac{\kappa_{\rm T}}{0.34 \, {\rm cm}^2 \, {\rm g}^{-1}}\right)^{-1} \ M_\odot \, {\rm yr}^{-1} \ .
\end{equation}
\par
A NS accreting mass can also accrete angular momentum and be spun up.
The accretion torque on a NS can be approximated as \citep[e.g.,][]{bildsten_gravitational_1998} 
\begin{equation} \label{eq:nacc}
N_{\rm acc} \approx {\dot M}\sqrt{G M_{\rm NS} R_{\rm eq}} \ ,
\end{equation}
where $M_{\rm NS}$ is the mass of the NS and $R_{\rm eq}$ is the radius of the NS at its equator.
If we assume torque balance, so that GW radiation losses equal the accretion torque (discussed in more detail in \S~\ref{sec:gw}), the estimated strain amplitude from the accreting neutron star is
\begin{equation}
h_0 \approx 1.8{\times}10^{-26} \left(\frac{D}{2.8 \, {\rm kpc}}\right)^{-1} \left(\frac{\dot M}{{\dot M}_{\rm Edd}}\right)^{1/2} \left(\frac{f_{\rm GW}}{600 \, {\rm Hz}}\right)^{-1/2} \ ,
\end{equation}
where we have scaled quantities to those appropriate for Sco X-1 and assumed $M_{\rm NS} = 1.33M_\odot$ and $R_{\rm eq} = 10{\rm\ km}$. Here $D$ is the distance to the LMXB and $f_{\rm GW}$ is the GW frequency.
\subsection{Bondi-Hoyle-Lyttleton accretion}
The accretion flow during CE inspiral can be parametrized using the theory of BHL accretion \citep{hoyle_effect_1939,bondi_mechanism_1944, edgar_review_2004}. The characteristic length scale for accretion onto an object of mass $M$ moving at a speed $v_\infty$ relative to a uniform medium with density $\rho_\infty$ and sound speed $c_{\rm s}$ is given by
\begin{equation}
R_{\rm BHL} \equiv \frac{2 G M}{v_\infty^2 + c_{\rm s}^2} \ .
\end{equation}
If, following MR15, we neglect the sound speed, then the characteristic length scale is the Hoyle-Lyttleton (HL) accretion radius, $R_{\rm HL}$ \citep{hoyle_effect_1939}.
We define the following parameters:
\begin{subequations}
\begin{align}
R_{\rm a} &\equiv R_{\rm HL} = \frac{2 G M_{\rm NS}}{v_\infty^2} \ , \\
\tau_{\rm a} &\equiv \frac{R_{\rm a}}{v_\infty} \ , \\
{\cal M}_\infty &\equiv \frac{v_\infty}{c_{\rm s}} \ .
\end{align}
\end{subequations}
Here $R_{\rm a}$ is the accretion radius of the NS, $\tau_{\rm a}$ is the inspiral accretion timescale, and ${\cal M}_\infty$, is the local upstream Mach number. The HL accretion rate is then
\begin{equation} \label{eq:hl_mdot}
{\dot M}_{\rm HL} = \pi R_{\rm a}^2 \rho_\infty v_\infty \ .
\end{equation}
Assuming this form for the accretion rate, the gravitational drag force \citep[e.g.,][]{iben_common_1993} and corresponding energy loss rate experienced by the inspiraling companion can be estimated using
\begin{subequations} \label{eq:hl_drag}
\begin{align}
F_{\rm d,HL} &= \pi R_{\rm a}^2 \rho_\infty v_\infty^2 = {\dot M}_{\rm HL} v_\infty \ , \\
{\dot E}_{\rm HL} &= \pi R_{\rm a}^2 \rho_\infty v_\infty^3 = {\dot M}_{\rm HL} v_\infty^2 = F_{\rm d,HL} v_\infty \ .
\end{align}
\end{subequations}
The actual accretion rate, drag force, and energy loss rate are reduced relative to the HL rates during CE inspiral as discussed in \S~\ref{sub:mr15}.
\subsection{Hyper-critical accretion during CE inspiral}
In hyper-critical accretion, thermal neutrino cooling is sufficient to allow the accretion rate to exceed the Eddington limit \citep{houck_steady_1991}.
Once the NS surface temperature exceeds $\sim 1 \  {\rm MeV}$, photons can be trapped in the flow and a neutrino-cooling layer is expected to form outside the NS surface.
The photon-trapping radius is approximately
\begin{eqnarray}
R_{\rm tr} &=&\frac{{\dot M}\kappa_{\rm T}}{4 \pi c} \\
\nonumber
&\approx& 5.7{\times}10^{13} \left(\frac{\dot M}{M_\odot \, {\rm yr}^{-1}}\right)\left(\frac{\kappa_{\rm T}}{0.34 \ {\rm cm}^2 \, {\rm g}^{-1}}\right) \ {\rm cm} \ .
\end{eqnarray}
The shock radius is approximately
\begin{equation}
R_{\rm sh} \approx 1.6{\times}10^8 \left(\frac{\dot M}{M_\odot \, {\rm yr}^{-1}}\right)^{-0.37} \ {\rm cm} \ .
\end{equation}
The general criterion for super-Eddington accretion to occur via efficient thermal neutrino cooling is $R_{\rm tr} > R_{\rm sh}$, which implies
\begin{eqnarray}
{\dot M}  \gtrsim {\dot M}_{\rm hyper} &\equiv& 8.9\times10^{-5}\left({{\kappa_{\rm T}}\over{0.34\ {\rm cm}^2\,{\rm g}^{-1}}}\right)^{-0.73}\\
\nonumber
&\sim& 10^4 {\dot M}_{\rm Edd}\ ,
\end{eqnarray}
where the last expression holds for typical NS masses.
In our model treatment, the transition from Eddington to hypercritical accretion occurs instantaneously.  
Physically, the density of the gas falling into the accretion radius of the NS is high enough such that temperatures rise to the point where thermal neutrino cooling will start to occur and allow for hyper-critical accretion.
\par
Farther away from the NS, transport of thermal radiation can efficiently cool the material that is gravitationally captured by the NS.
As a NS inspirals within the CE, it creates density wakes that provide additional gravitational drag.
Photon bubbles can travel faster through the less dense regions \citep[e.g.,][]{gammie_photon_1998} and transport thermal energy away from the accreting material more efficiently than advection and conduction.
Regions near the NS may also be unstable to the neutrino-bubble \citep{socrates_neutrino_2005} and photon-bubble instability, such that perturbations in the flow are seeded, grow exponentially, and saturate into shock trains \citep[e.g.,][]{turner_effects_2005,begelman_nonlinear_2006,turner_photon_2007}.
\par
As emphasized by MR15, a finite density gradient in the stellar envelope breaks the symmetry of HL accretion.
Material gravitationally captured within the accretion radius may not be accreted onto the inspiralling companion due to the flow's angular momentum.
The background density gradient can be parametrized using the non-dimensional quantity 
\begin{equation}
\epsilon_\rho \equiv \frac{R_{\rm a}}{H_\rho} = - \frac{R_{\rm a}}{\rho} \frac{\rm d\rho}{{\rm d}r} \ ,
\end{equation}
where $\rho$ is the density of the background medium and $H_\rho$ is the density scale height.
MR15 found that the accretion rate decreases relative to ${\dot M}_{\rm HL}$ with steeper density gradients.
(We describe the MR15 model in more detail in \S~\ref{sec:methods}.)
This result suggests that NSs can survive the CE phase since they accrete only a small fraction of their initial mass during CE inspiral.
The actual accretion rate ${\dot M}$ during inspiral appears to satisfy ${\dot M}_{\rm Edd} < {\dot M}_{\rm hyper} < {\dot M} < {\dot M}_{\rm BHL} < {\dot M}_{\rm HL}$. 
Hence, using ${\dot M}_{\rm HL}$ for the accretion rate should provide an upper limit on the GW emission due to accretion during CE inspiral, assuming torque balance. 
Since the accretion rate suppression in the MR15 model can be significant, we use MR15 instead of the HL rate in what follows.
\section{Gravitational Waves from Accreting Neutron Stars} \label{sec:gw}
In the weak-field limit, the GW luminosity from a spinning NS with a finite quadrupole moment $Q$ is approximately
\begin{equation} \label{eq:lgw}
L_{\rm GW} = \frac{32}{5} \frac{G \Omega^6 Q^2}{c^5} \ ,
\end{equation}
where $\Omega$ is the spin angular frequency of the NS.
The frequency of the emitted GWs is twice the spin frequency: $f_{\rm GW} = 2 f_{\rm spin} = \Omega/\pi$.
The corresponding torque exerted from the GW emission of angular momentum is
\begin{equation} \label{eq:ngw}
N_{\rm GW} = -{{L_{\rm GW}}\over{\Omega}} \ .
\end{equation}
For optimally oriented GW sources, the GW strain amplitude $h_0$ measured by a distant GW detector is 
\begin{equation} \label{eq:strain}
h_0 = \frac{2 G \Omega^2}{c^4 D} Q = \frac{2 G \Omega^2}{c^4 D} \varepsilon I \ ,
\end{equation}
where $D$ is the distance between the source and the detector, $\varepsilon$ is the quadrupole ellipticity, and $I = \frac{1}{5} M_{\rm NS} \left(R_z^2 + R_e^2\right)$ is the moment of inertia of the NS.
If the accretion torque is greater than the GW torque, then the NS spins up according to
\begin{equation} \label{eq:spinup}
I \dot{\Omega} = N_{\rm acc} + N_{\rm GW} \ .
\end{equation}
We will find that the duration of the NS spin-up depends on the accretion torque, such that the frequency may be relatively constant depending on the coherence time used for the integration.
\subsection{NS Mountains} \label{ssec:mount}
NS surface deformations away from axisymmetry, i.e., mountains, produce a time-varying quadrupole moment and thus GW emission.
There are two types of NS mountains that may be present as the NS accretes during CE evolution: thermal mountains and magnetic mountains. 
\cite{bildsten_gravitational_1998} first described how electron captures in the crust of Eddington-accreting NSs may generate density asymmetries on the NS surface if a transverse temperature gradient is present, i.e., either from compositional variation, asymmetric local accretion flow, buried magnetic fields, etc., and thus lead to GW emission.  
The typical temperature of the crust of Eddington-accreting NSs is of order $10^{8} \ {\rm K}$ such that the crust remains solid.  
The maximum quadrupole moment of such thermal mountains depends on the properties of the NS crust where the crust ellipticity can be approximated as \citep{glampedakis_gravitational_2017}
\begin{equation}
\varepsilon_{\rm th} \lesssim \frac{\mu_{\rm cr} \sigma_{\rm br} V_{\rm cr}}{G M_{\rm NS}^2/R_{\rm NS}} \sim 10^{-5} \left(\frac{\sigma_{\rm br}}{0.1}\right) \ ,
\end{equation}
where $\mu_{\rm cr}$ is the shear modulus of the crust, $V_{\rm cr}$ is the volume of the crust, and $\sigma_{\rm br}$ is the crustal breaking strain. 
For magnetic mountains on NSs with magnetar-level field strengths, $B\sim 10^{15} \, {\rm G}$, the ellipticity may take on values of $\varepsilon_{\rm mag} \sim 10^{-6}-10^{-5}$. 
Our work is agnostic to the mechanism by which mountains are produced and only requires that such quadrupole moments are achievable.  
Several previous works have estimated what values of the maximum quadrupole moment may be sustained by the NS crust.  
\cite{haskell_mountains_2006} applied the formalism presented by \cite{ushomirsky_deformations_2000} to both accreted and non-accreted NS crusts and determined the maximum quadrupole moment to be $Q_{\rm max} \approx 2{\times}10^{39} \ {\rm g} \, {\rm cm}^2$. 
\cite{horowitz_breaking_2009} performed 3D molecular dynamics simulations of local regions of NS crust to obtain estimated for the breaking strain.  
With their estimate for the breaking strain, \cite{horowitz_breaking_2009} obtain a maximum NS crust ellipticity of $\varepsilon \approx 4{\times}10^{-6}$.
With this bound on the ellipticity, we thus use a maximum quadrupole moment of $Q_{\rm max} = 4{\times}10^{39} \ {\rm g} \, {\rm cm}^2$ for this work.  
With this upper limit on the maximum quadrupole moment, the strain amplitudes we estimate in \S~\ref{sec:res} should be regarded as upper limits.  
\par
When the accretion torque approximately balances the GW radiation torque $(N_{\rm GW} + N_{\rm acc} \approx 0)$, the quadrupole moment can be expressed as a function of the accretion rate by using \eqref{eq:nacc} and \eqref{eq:ngw} to obtain
\begin{equation} \label{eq:quad}
Q_{\rm tb} = \left(\frac{5}{32} \frac{c^5}{G \Omega^5} {\dot M} \sqrt{G M_{\rm NS} R_{\rm eq}} \right)^{1/2} \ .
\end{equation}
Since the mass and accretion rate change slowly compared to the GW period, the implicit time dependence they introduce into $Q$ modulates the GW strain amplitude as the NS inspirals within the CE.
\par
We treat our model NS undergoing CE evolution as beginning in torque balance and spinning up if the required quadrupole moment to maintain torque balance is greater than the maximum quadrupole moment that the NS crust can sustain.
At hyper-critical accretion rates temperatures are of order MeV such that the  the solid NS crust melts into liquid and erases surface deformations.
As long as there exists solid NS crust, thermal mountains and/or magnetic mountains may still be formed before they melt away.
Our treatment of the crust-melting process is described in \S~\ref{ssec:quad}.
For the magnetar scenario, the crust remains deformed by the global magnetic field as it is melting.
For the thermal-mountain scenario, the electron-capture rates just need to vary enough with temperature in order to produce the required quadrupole moment.
We include a derivation of the required electron-capture rate temperature dependence in Appendix \ref{appa}.
At accretion rates relevant to the fallback scenario, the accretion torque is large enough such that the NS is spun up to secular instability, which excites $r$-modes that produce GWs.
In \S~\ref{sec:res} we will show that accreting NSs in the CE case do not spin up to secular instability.
%
\section{Methods} \label{sec:methods}
Here we describe how we use the considerations in the preceding sections to predict the GW signal produced by NS companions in CEs with supergiant primaries. 
The steps we take are:
\begin{enumerate}[(1)]
\item Construct a model for the primary's density and temperature structure at the time of CE inspiral. 
We use MESA for this task.
\item Construct a model for the accretion-rate evolution. 
We use the MR15 model for the accretion rate as a function of the background density gradient.
\item Construct a model for the orbital evolution.  
We take the orbital decay to be driven entirely by local gravitational drag as computed using the MR15 model.
\item Integrate the orbital evolution equation until the energy dissipated in the envelope is equal to the binding energy of the envelope for a given choice of $\alpha_{\rm CE}$.  
\item Construct a model for GW emission from the spinning NS. 
The strain amplitude $h_0$ is computed using the NS spin frequency $\Omega$ and the NS quadrupole moment $Q$.
We describe the details of our model in \S~\ref{ssec:spin} and \S~\ref{ssec:quad}.
\end{enumerate}
\subsection{Primary stellar model}
\begin{figure}
\centering
\includegraphics[width=\columnwidth]{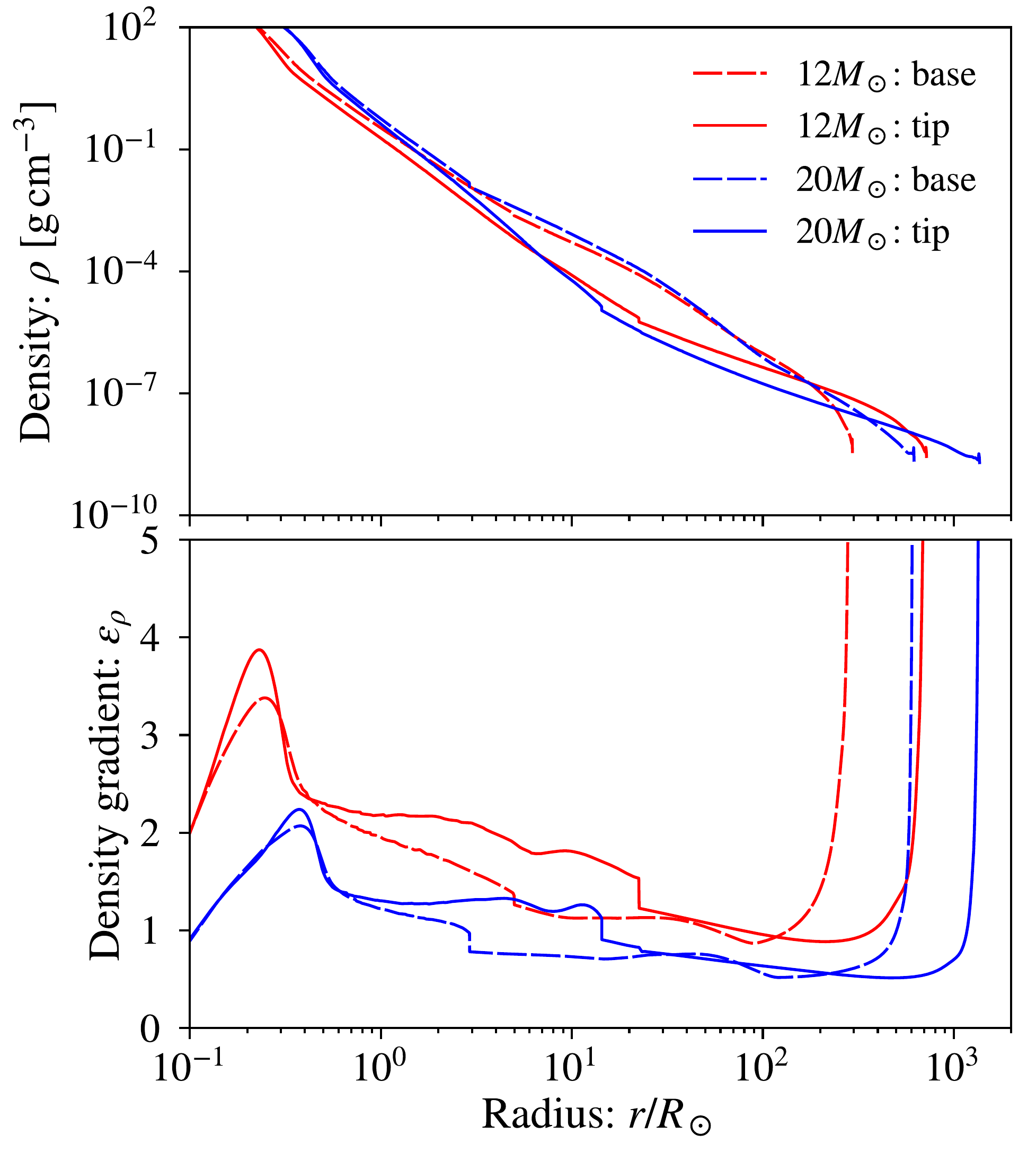}
\caption{\label{fig:eps}Top panel: density profiles for $12 M_\odot$ and $20 M_\odot$ MESA stellar models at the base and tip of the RGB.
Bottom planel: non-dimensional density gradients for the same models.
}
\end{figure}

We compute the single-star evolution of a $12 M_\odot$ and a $20M_\odot$ primary with solar metallicity using the MESA stellar evolution code version 8845 \citep {paxton_modules_2011,paxton_modules_2013,paxton_modules_2015}. 
We consider CE phases occurring at two different evolutionary stages: at the base and the tip of the red giant branch (RGB). Due to mass loss, the precise donor star masses used are slighly smaller than their zero-age main sequence (ZAMS) values.\footnote{MESA inlists and the model profiles used in this paper will be made available through the MESA Marketplace at \url{http://cococubed.asu.edu/mesa_market/}.}
During the inspiral the primary's structure is taken to be fixed.
As pointed out by MR15, this approximation is most reasonable for large binary mass ratios.
The density profiles and non-dimensional density gradients of our stellar models are shown in Figure \ref{fig:eps}.
A density inversion appears near the surface of the stellar models because of the interaction of the iron opacity bump with the nearly-Eddington radiative flux in the envelope. 
This produces a spike in $\epsilon_\rho$. 
3D radiation hydrodynamics simulations \citep{jiang_local_2015} show that in low-density regions such inversions develop a porous structure that allows radiative and convective energy transport to compete, expanding the envelope and partially eliminating the inversion. 
In our calculations, the presence of the inversion does not matter significantly for the GW prediction. 
Given the present uncertainty regarding how to properly treat such inversions in 1D stellar evolution codes we have chosen to leave the inversion unmodified in this paper.
\subsection{Accretion rate model} \label{sub:mr15}
\begin{figure*} 
\centering
\includegraphics[width=0.9\textwidth]{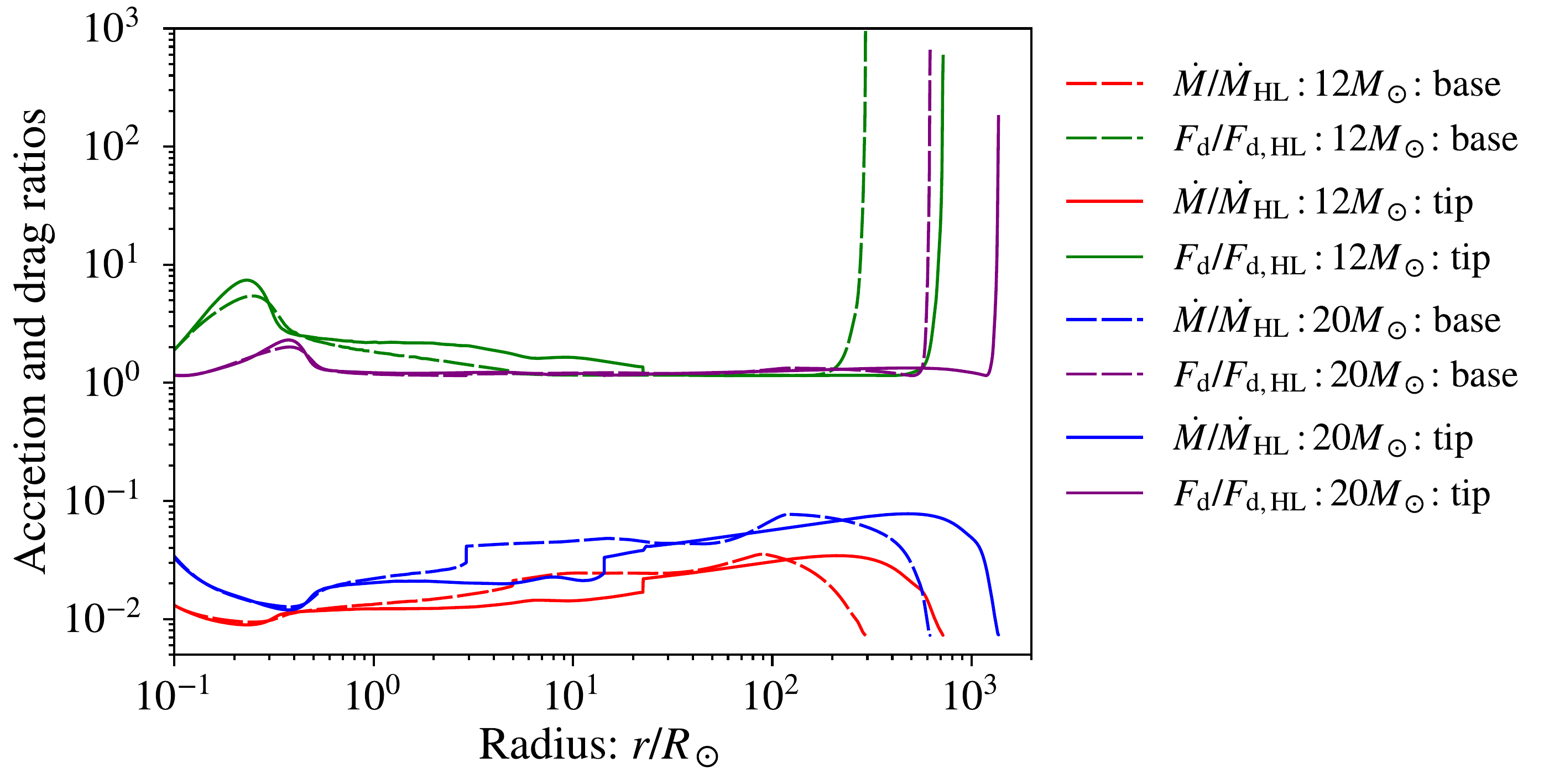}
\caption{\label{fig:ratios}
Drag and accretion rate (relative to HL) given the envelope structure of our $12 M_\odot$ and $20M_\odot$ stars.
}
\end{figure*}

To estimate the NS accretion rate at a given radius within the primary we use the HL rate expressions (Equations \eqref{eq:hl_mdot} and \eqref{eq:hl_drag}) modified using the model of MR15. 
MR15 performed simulations of accretion onto a spherical sink region with a background density gradient as a model for the local accretion dynamics during CE evolution. 
They approximate the gravitational drag obtained in their simulations using
\begin{equation} \label{eq:fd_norm}
\frac{F_{\rm d,MR15}(\epsilon_\rho)}{F_{\rm d,HL}} \approx {\rm f}_1 + {\rm f}_2 \epsilon_\rho + {\rm f}_3 \epsilon_\rho^2 \ ,
\end{equation}
where the coefficients ${\rm f}_i$ ($i = 1,2,3$) are given by
\begin{subequations}
\begin{align}
{\rm f}_1 &= 1.91791946 \ , \\
{\rm f}_2 &= -1.52814698 \ , \\
{\rm f}_3 &= 0.75992092 \ .
\end{align}
\end{subequations}
The accretion rate in their simulations is approximately 
\begin{equation} \label{eq:mdot_norm}
\log \left(\frac{{\dot M}_{\rm MR15} (\epsilon_\rho)}{{\dot M}_{\rm HL}}\right) \approx 
\mu_1 + \frac{\mu_2}{1 + \mu_3 \epsilon_\rho + \mu_4 \epsilon_\rho^2} \ , 
\end{equation} 
with accretion-rate coefficients $\mu_i$ ($i = 1,2,3,4$) given by
\begin{subequations}
\begin{align}
\mu_1 &= -2.14034214 \ , \\
\mu_2 &= 1.94694764 \ , \\
\mu_3 &= 1.19007536 \ , \\
\mu_4 &= 1.05762477 \ .
\end{align}
\end{subequations}
We plot the accretion and drag ratios computed using Equations \eqref{eq:fd_norm} and \eqref{eq:mdot_norm} and the density gradients from our MESA stellar models in Figure \ref{fig:ratios}.
Note that in the region of the density inversion the drag is significantly enhanced while the accretion rate is significantly diminished relative to the HL prediction. 
Hence including the inversion in the primary model causes the NS to pass through this region rapidly without emitting much GW energy. 
In some respects this mimics the effect of the initial dynamical plunge seen in 3D CE simulations, which we do not explicitly include.
\subsection{Accretion rate parametrization}
Accretion onto the NS during the CE phase is a fairly uncertain process, and the wind-tunnel model for the orbital evolution may not work well in our case since the primary star structure may respond to the orbiting NS. 
To consider the possibility that the accretion rate is diminished even further below the MR15 model, we parameterize the accretion rate using an efficiency factor $\eta \le 1$.
Since unstable cooling solutions exist for ${\dot M}_{\rm Edd} < {\dot M} < {\dot M}_{\rm hyper}$, the accretion rate we use is
\begin{equation} \label{eq:hrh17}
{\dot M} = 
  \begin{cases} 
   \eta {\dot M}_{\rm MR15} & \eta{\dot M}_{\rm MR15} \ge {\dot M}_{\rm hyper}{\rm\ or\ }\le {\dot M}_{\rm Edd} \ , \\
   {\dot M}_{\rm Edd}       & {\dot M}_{\rm Edd} < \eta {\dot M}_{\rm MR15} < {\dot M}_{\rm hyper} \ .
  \end{cases}
\end{equation}
The case ${\dot M} < {\dot M}_{\rm Edd}$ is implicitly allowed in our model, but in our integrations we find that the accretion rate is always {\bf $\ge \dot M_{\rm Edd}$.}  
The corresponding drag force is then
\begin{equation} \label{eq:dragactual}
F_{\rm d} = \eta F_{\rm d,MR15} \ ,
\end{equation}
where even for Eddington-limited accretion rates, momentum dissipation occurs within the accretion radius of the NS which contributes to the drag force.
\subsection{Orbital evolution model} 
We consider the same orbital evolution model as MR15. 
In this model the inspiral is driven by local gravitational drag within the envelope such that the orbital decay rate, ${\dot a}$, obeys
\begin{equation} \label{eq:adot}
\dot{a} = \dot{E} \frac{da}{dE_{\rm orb}} \ ,
\end{equation}
where ${\dot E}$ is the energy dissipated by gravitational drag,
\begin{equation}
{\dot E} = - F_{\rm d} (\epsilon_\rho) v_\infty \ .
\end{equation}
The gravitational drag force $F_{\rm d}(\epsilon_\rho)$ is obtained from Equation \eqref{eq:dragactual}. 
Assuming a nearly circular orbit, the inspiral velocity $v_\infty$ obeys
\begin{equation} \label{eq:vinf}
v_\infty^2 = \frac{G (M_{\rm NS} + m(a))}{a} \ ,
\end{equation}
where $m(a) = \int_0^a 4 \pi \rho(r) r^2 \, {\rm d}r$ is the envelope mass enclosed within the orbital separation. 
The orbital energy $E_{\rm orb}$ is
\begin{equation}
E_{\rm orb} = -\frac{G M_{\rm NS}m(a)}{2a} \ .
\end{equation}
We follow the evolution of the accretion rate, $\dot{M}$, and the orbital separation, $a$, by integrating Equation~\eqref{eq:adot}. 
Following MR15, we terminate the orbital evolution when the dissipated energy begins to exceed the binding energy of the envelope, 
\begin{equation} \label{eq:stop}
\Delta E_{\rm orb} \ge E_{\rm bind} \ ,
\end{equation}
where we take $\alpha_{\rm CE} = 1$ and where the binding energy is
\begin{equation}
E_{\rm bind} = \int_{m(a)}^M \left(u - \frac{G m(a)}{r} \right) \, {\rm d}m \ .
\end{equation}
Here $u$ is the specific internal energy of the envelope gas.
This criterion may not correspond to the true end of the CE phase.
\par
From the strain amplitude and/or its time derivative, it may be possible to constrain the orbital decay rate of the NS.
We thus define the quantity
\begin{equation}
\xi \equiv \frac{\tau_{\rm ins}}{\tau_{\rm orb}} \ ,
\end{equation} 
where the inspiral timescale $\tau_{\rm ins}$ and the orbital timescale $\tau_{\rm orb}$ respectively obey
\begin{subequations}
\begin{align}
\tau_{\rm ins} &= \frac{a}{|\dot a|} \ , \\
\tau_{\rm orb} &= \frac{2\pi}{\sqrt{G(M_{\rm NS} + m(a))}} a^{3/2}\ ,
\end{align}
\end{subequations}
and ${\dot a}$ obeys Equation~\eqref{eq:adot}.
We use the $\xi$ parameter to characterize how the orbital decay of the NS evolves within the CE and describe its features as follows.
The limit of $\xi \gg 1$ corresponds to a slow migration, where the inspiral occurs over many orbital periods.  
The range $\xi \sim 1$ corresponds to a merger within an orbital period, i.e., a moderate inspiral.
The limit of $\xi \ll 1$ corresponds to a merger on a timescale much shorter than the orbital period, i.e., a rapid inspiral.
In terms of CE evolution, the plunge-in phase may be characterized by $\xi \ll 1$ increasing to $\xi \sim 1$, followed by the self-regulated phase, where $\xi \gg 1$. 
Since we are treating the CE inspiral as quasi-circular, cases where $\xi \ll 1$ indicate a breakdown in the validity of this assumption.
\subsection{Spinning NS model} \label{ssec:spin}
To self-consistently determine the equatorial radius given the NS spin, we adopt the same approach as PT12.
We treat the NS as an axisymmetric Maclaurin spheroid and take any internal viscous dissipation to be negligible since it operates on timescales several orders of magnitude longer than dissipation via GW radiation.
The GW emission then reacts immediately to changes in the accretion rate.
\par
\begin{figure}
\centering
\includegraphics[width=\columnwidth]{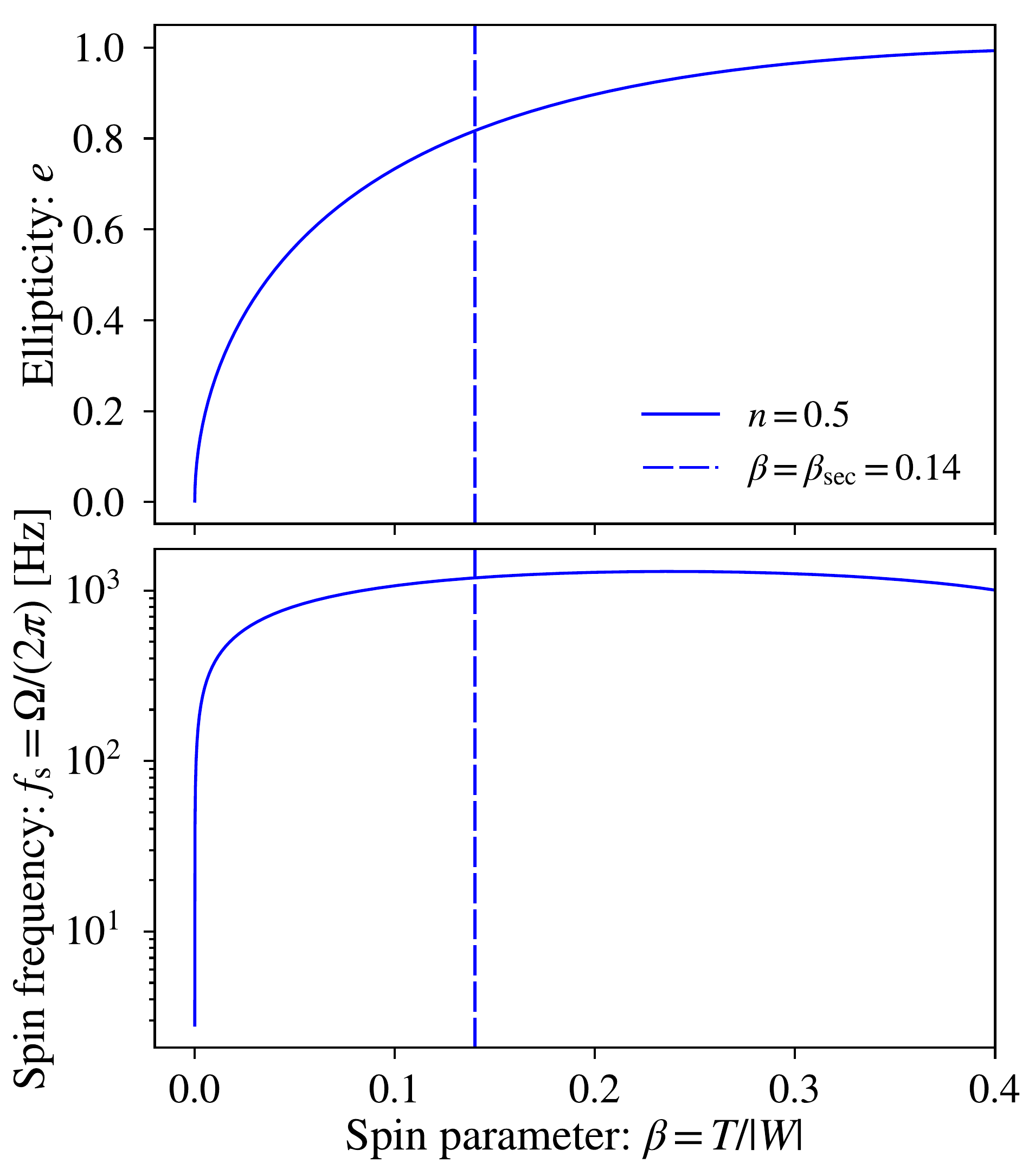}
\caption{\label{fig:spin} Solutions of the Maclaurin spheroid model for $n = 0.5$.
The green dashed line indicates the threshold of secular instability.
Top panel: ellipticity vs. spin parameter.  
Bottom panel: spin frequency vs. spin parameter.
}
\end{figure}

The NS spin is parametrized using the spin parameter $\beta = T/|W|$, where $T$ is the rotational kinetic energy and $W$ is the gravitational potential energy.
From the Maclaurin spheroid model, the spin parameter can be expressed in terms of the NS ellipticity as
\begin{equation}
\beta = \frac{3}{2 e^2} \left[1 - \frac{e(1 - e^2)^{1/2}}{\sin^{-1} e}\right] - 1 \,.
\end{equation}
%
%
%
The NS spin obeys
\begin{equation}
\Omega^2 = \frac{2 \pi G {\bar \rho}}{q_n} \left[\frac{(1 - e^2)^{1/2}}{e^3} (3-2e^2) \sin^{-1}e - \frac{3(1 - e^2)}{e^2}\right] \ ,
\end{equation}
where ${\bar \rho} = 3 M / 4 \pi R_0^3$ is the average density, $R_0$ is the radius of the NS with zero rotation, and
\begin{equation}
q_n = (1 - n/5) \kappa_n \ ,
\end{equation}
with $n$ as the polytropic index and $\kappa_n$ as a constant of order unity \citep[see Table 1 in][]{lai_ellipsoidal_1993}.
We take $n=0.5$, which is stable against mass shedding. 
From the ellipticity, the NS's equatorial radius is
\begin{equation}
R_{\rm eq} = \frac{\bar R}{(1-e^2)^{1/6}} \ ,
\end{equation}
where ${\bar R}$ is the mean radius of the NS given by
\begin{equation}
\bar R = R_0 \left[\frac{\sin^{-1}e}{e} (1 - e^2)^{1/6}(1- \beta)\right]^{-n/(3-n)} \ .
\end{equation}
We take the spin parameter to be bounded by $\beta_{\rm sec}$, so that GW emission prevents the NS from becoming secularly unstable.
We plot the ellipticity and spin frequency, $f_{\rm s} = \Omega/(2\pi)$, as a function of the spin parameter in Figure \ref{fig:spin}.
%
\subsection{Quadrupole-moment model and strain amplitude} \label{ssec:quad} 
The NS crust ellipticity is the source of GWs in our model of NS-CE evolution.  
The temperature of the NS crust while accreting at the Eddington rate are $\sim 10^{8} \ {\rm K}$ \citep[e.g.,][]{bildsten_gravitational_1998}. 
At MeV temperatures relevant to hyper-critical accretion, the outer crust melts into ocean, which erases surface deformations that were previously present.  
The lower layers may still have surface deformations and continue to generate GWs before melting away.  
Once the entire crust melts, then there is no GW emission from the NS for the rest of the CE evolution.  
In this scenario, we expect there to be GW emission as long as the crust-melting timescale is longer than the CE inspiral timescale.  
Physically, the local crust-melting rate depends on a number of factors including the composition, density, etc., which are uncertain.   
As a conservative assumption, we take the accretion rate as a bound on the crust-melting rate, which sets the crust-melting timescale to
\begin{equation}
\tau_{\rm melt} \sim \frac{M_{\rm cr}}{\dot M} \ ,  
\end{equation}
A lower crust-melting rate would be advantageous as it would allow for the NS crust to remain present on a longer timescale, thus allowing for more GW emission to occur during the CE inspiral. 
We approximate the crust to be close to constant density such that the ellipticity of the crust is a function of the crust mass that is present, thus setting the maximum quadrupole moment to
\begin{equation}
Q_{\rm max} = 4{\times}10^{39} \left(\frac{M_{\rm acc}}{M_{\rm cr}}\right) \ {\rm g} \, {\rm cm}^2 = 4{\times}10^{39} \left(\frac{M_{\rm acc}}{0.05 M_\odot} \right) \ {\rm g} \, {\rm cm}^2\ ,
\end{equation}
where $M_{\rm acc}$ is the amount of mass the NS accretes during CE inspiral and where we have taken the initial crust mass to be $M_{\rm cr} = 0.05 M_\odot$ \citep{barkov_recycling_2011}.
The model for the quadrupole moment is summarized as:
\begin{equation} \label{eq:hrh17}
Q = \min(Q_{\rm max},Q_{\rm tb}) \ ,
\end{equation}
where $Q_{\rm tb}$ is the quadrupole moment obtained from torque balance, Eq.~\eqref{eq:quad}, as described in \S~\ref{ssec:mount}.
With the above prescription, the NS crust sustains a maximum quadrupole moment and spins up according to Eq.~\eqref{eq:spinup} if $Q_{\rm tb} < Q_{\rm max}$.
If $Q_{\rm tb} > Q_{\rm max}$, then the NS maintains torque balance. 
With the spin frequency $\Omega$ and the quadrupole moment $Q$, the strain amplitude $h_0$ is computed using Eq.~\eqref{eq:strain}. 
%
%
\section{Results} \label{sec:res}
\subsection{Model parameters and initial conditions}
The parameter values we use for the spinning NS are $M_{\rm NS} = 1.33 \, M_\odot$ and $R_{\rm NS}= R_0 = 10\ {\rm km}$.
We take the NS to have a spin frequency of $\Omega/(2\pi) = 250 \ {\rm Hz}$ with the corresponding ellipticity and spin period being $e_0 = 0.177$ and $P_0 = 4 \ {\rm ms}$, respectively.
In order to factor out the distance to the source, we will present the strain amplitude relative to its Eddington-accreting NS in torque balance spinning at the initial rotation rate, $P_0$.
The initial separation is taken to be $a_{\rm i} = R_1$, where $R_1$ is the radius of the primary.
We summarize the parameter values in Table \ref{tab:model}.
\begin{table}
\centering
\caption{Model parameters.}
\label{tab:model}
\begin{tabular}{lcccc}
\hline 
Model  & $\eta$ & $ M_{\rm ZAMS}/M_{\rm actual} (M_\odot)$ &  Stage \\ \hline
M1 & 1.0 & 12/11.8 & RGB base \\ 
M2 & 0.1 & 12/11.8 & RGB base \\ 
M3 & 1.0 & 12/11.7 & RGB tip \\
M4 & 0.1 & 12/11.7 & RGB tip \\ 
M5 & 1.0 & 20/19.2 & RGB base \\ 
M6 & 0.1 & 20/19.2 & RGB base \\ 
M7 & 1.0 & 20/19.1 & RGB tip \\
M8 & 0.1 & 20/19.1 & RGB tip \\ \hline 
\end{tabular}
\end{table}
\subsection{Orbital evolution}
\begin{figure*} 
\centering
\includegraphics[width=\textwidth]{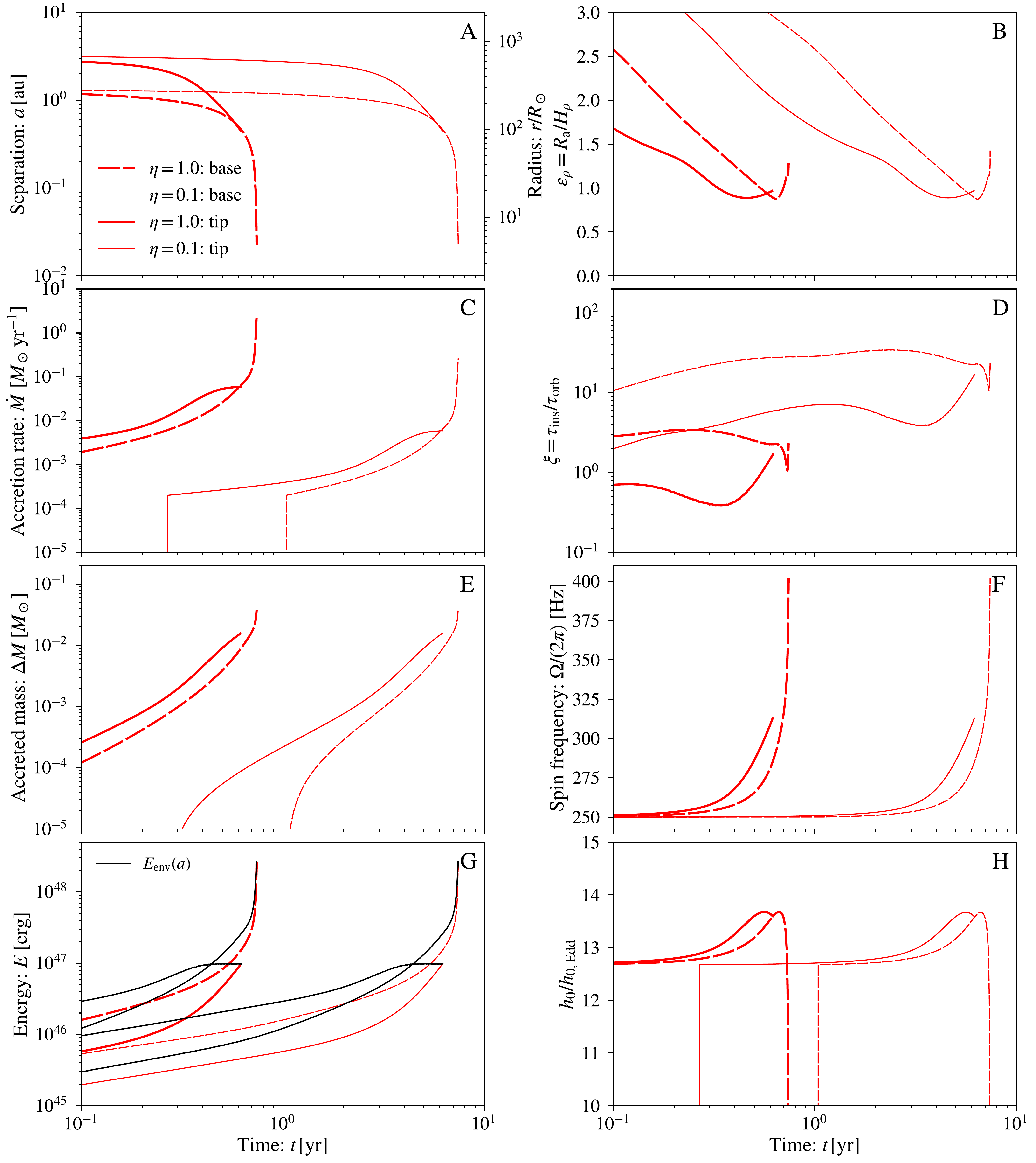}
\caption{
Orbital evolution curves for the $12 M_\odot$ stellar model.
Panel A: orbital separation between the NS and the primary's core, in units of ${\rm au}$ and $R_\odot$.
Panel B: non-dimensional density gradient the NS experiences during the inspiral.  
Panel C: NS accretion rate in units of $M_\odot \ {\rm yr}^{-1}$.  
Panel D: inspiral parameter.
Panel E: total accreted mass in $M_\odot$.
Panel F: spin frequency.
Panel G: dissipated orbital energy.  The binding energy of each stellar model is plotted as black curves.  The orbital integration is terminated according to Equation~\eqref{eq:stop}.  
Panel H: GW strain amplitude ratio relative to a NS accreting at its Eddington limit.
The thick and thin curves correspond to $\eta = 1$ and $\eta = 0.1$, respectively.
The dashed and solid curves correspond to our models at the base and tip of the RGB, respectively.}
\label{fig:orb1}
\end{figure*}

\begin{figure*} 
\centering
\includegraphics[width=\textwidth]{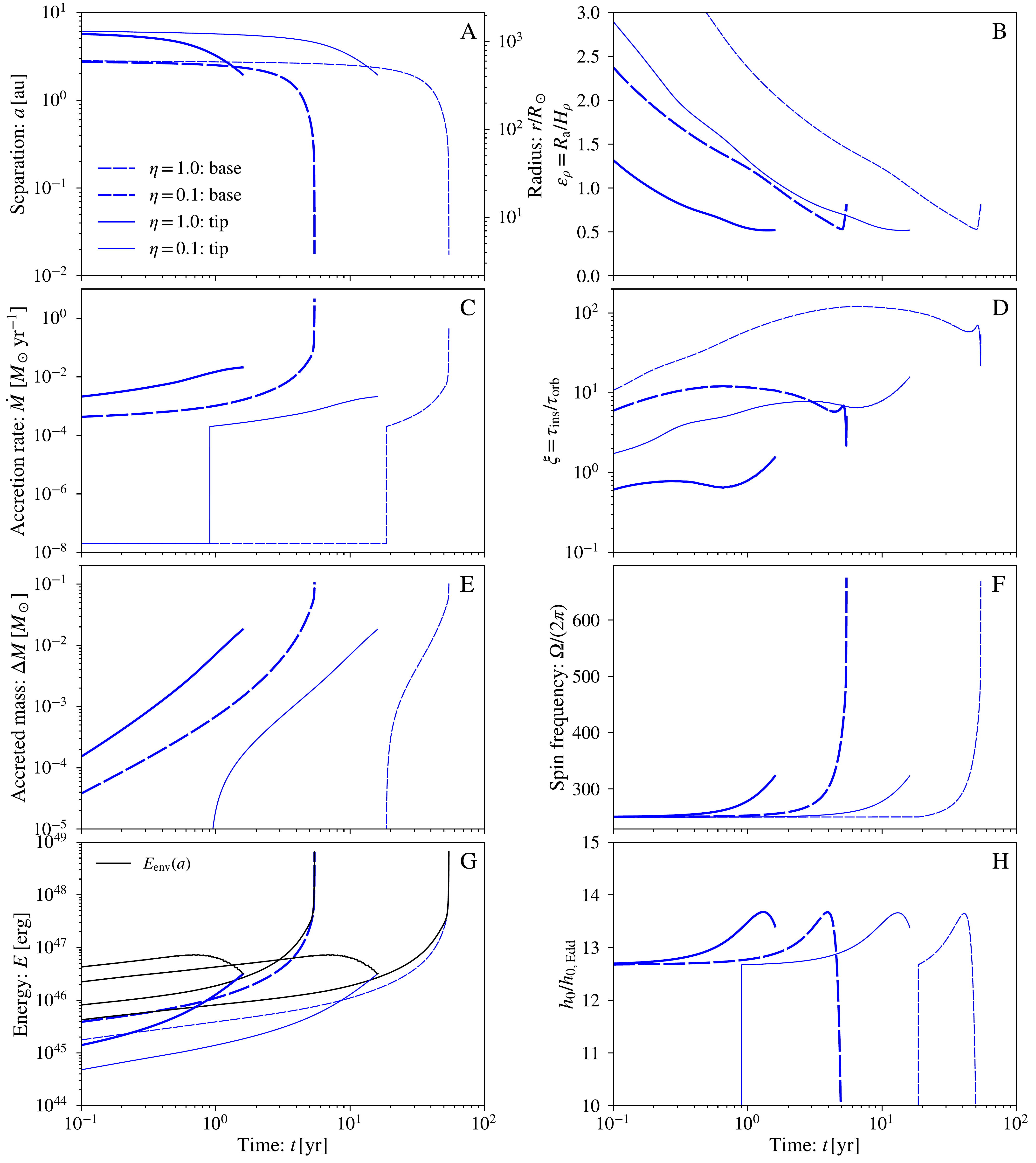}
\caption{
Same quantities as Figure~\ref{fig:orb1} for the $20 M_\odot$ model.
}
\label{fig:orb2}
\end{figure*}

Given the NS parameters and the primary structure, we determine the orbital separation, accretion rate, and strain amplitude as functions of time by integrating Equation \eqref{eq:adot}. 
The resulting evolution for the $12 M_\odot$ and $20 M_\odot$ models is shown in Figures~\ref{fig:orb1} and \ref{fig:orb2}.

In all cases we find that the accreted mass (panel E in Figures \ref{fig:orb1} and \ref{fig:orb2}) is less than $\sim 0.1M_\odot$ at the end of the integration, consistent with MR15. Thus the NS does not accrete enough mass to collapse into a BH. However, for both initial masses, the inspiral behavior (panel A) is strikingly different for stars at different evolutionary stages. For donors at the tip of the RGB, the NS appears to end the inspiral at a finite separation, with $\xi$ increasing. However, for donors at the base of the RGB, the NS undergoes a rapid plunge upon reaching the termination criterion, suggesting that these systems will merge and form Thorne-Zytkow objects. In these cases the accretion rate increases rapidly at the end of the inspiral. We have tested the sensitivity of this result to numerical effects by carrying out the integrations with 1/10 the timestep and integrating past the termination criterion, finding that the behavior is unchanged.
\par
Careful comparison with the model profiles (Figure~\ref{fig:eps}) suggests that the more evolved stars avoid merger because they reach the termination criterion before the NS reaches the hydrogen burning shell, where $\epsilon_\rho$ displays a discontinuity. 
Although reaching a burning shell should violate the assumption that the donor profile is static (and the final plunge certainly violates the quasi-circular orbit assumption), the response of the star at this point might be to expand and thus further increase the drag. 
However, note from panel D that except for the more evolved cases with $\eta = 1.0$ and the very end of the merger cases (if run past the termination criterion), generally $\xi > 1$. 
This suggests that the quasi-circular approximation is at least self-consistent for these cases.
\par
The strain amplitude (panel H) is typically louder than an Eddington-limited NS, particularly for the merger cases, implying that NS-CE systems can potentially be louder sources of GWs than LMXBs.
This also implies that NS-CE systems may potentially be detected out to greater distances than Galactic LMXBs, which we discuss in more detail in Section \ref{sub:phase}.
Since the accretion rate monotonically increases over the course of the inspiral (see panel C of Figures \ref{fig:orb1} and \ref{fig:orb2}), the strain amplitude also increases with time, but experiences a turn-over as a significant amount of crust has melted.   
For the more evolved stars, it levels off at $\sim 3-10\times$ its initial value, while for the less evolved stars, which merge, the strain amplitude increases by another factor of $\sim 3-10$ over a timescale of less than a year. 
The loudest NS-CE systems should therefore be the ones with more compact, less evolved donors. Merging and non-merging cases can be distinguished based on the sign of the second time derivative of the strain amplitude.
\par
Reducing the accretion efficiency $\eta$ below 1.0 can cause the accretion rate to fall within the region of unstable cooling solutions found by \cite{houck_steady_1991}.
This is seen in the $\eta = 0.1$ models.
The accretion rates during those periods are thus modulated by the Eddington limit, making the system no louder than a LMXB.
The gravitational drag decreases when the accretion rate is reduced, resulting in a longer inspiral timescale.
Thus the $\eta = 0.1$ models have an inspiral duration $\sim 10$ times longer than for $\eta=1$.
We tabulate the duration of the inspiral, the factor by which the initial orbital separation decreases, and the factor by which the strain amplitude increases in our models in Table \ref{tab:fac1}.
\begin{table}
\centering
\caption{\label{tab:fac1} Inspiral results.
}
\begin{tabular}{lcccc}
\hline
Model & $t_{\rm final} \ ({\rm yr})$ & $a_{\rm i}/a_{\rm f}$& $h_{0,\rm max}/h_{0,\rm i}$ & {\bf $\Omega_{\rm f}/\Omega_{\rm i}$} \\ \hline
M1 & $0.741$ & $59.9$ & {\bf $13.7$} & {\bf $1.61$} \\
M2 & $7.41$ & $61.5$ & {\bf $13.7$} & {\bf $1.61$} \\
M3 & $0.620$ & $7.47$ & {\bf $1.08$} & {\bf $1.25$} \\
M4 & $6.21$ & $7.54$ & {\bf $13.7$} & {\bf $1.25$} \\
M5 & $5.43$ & $161$ & {\bf $13.7$} & {\bf $2.70$} \\
M6 & $54.4$ & $165$ & {\bf $13.7$} & {\bf $2.68$} \\ 
M7 & $1.61$ & $3.28$ & {\bf $1.08$} & {\bf $1.30$} \\ 
M8 & $16.1$ & $3.29$ & {\bf $13.7$} & {\bf $1.29$} \\ \hline 
\end{tabular}
\end{table}
%
\subsection{Spin evolution}
The spin evolution is shown in Panel F of Figures~\ref{fig:orb1} and \ref{fig:orb2} and spin ratios and values are tabulated in Tables \ref{tab:fac1} and \ref{tab:fac2}, respectively.
The spin-up in our models suggest that the GW waveform is that of slow chirp, where both the GW amplitude and GW frequency increase over the course of the inspiral as long as there exists a significant amount of crust.
Smaller accretion torques produce a slower spin-up such that the waveform may be relatively continuous over an integration time of ${\cal O}(10 \, {\rm hr})$. 
Our models do not reach secular instability.
%
\section{Discussion} \label{sec:disc}
\subsection{Constraining the Orbital Separation}
\label{sec:orbsep}
The orbital separation at any point during CE evolution would be difficult to measure with electromagnetic observations due to the high optical depth of the envelope.
Direct detection of a GW signal from an accreting NS within the CE may provide the only way to constrain the orbital separation between the NS and the primary's core during the inspiral. 
However, since this constraint relies on knowledge of the primary mass and evolutionary state, the NS mass, and the distance to the system, pre-CE electromagnetic observations would be needed to obtain useful constraints. 
\par
\begin{figure}
\centering
\includegraphics[width=\columnwidth]{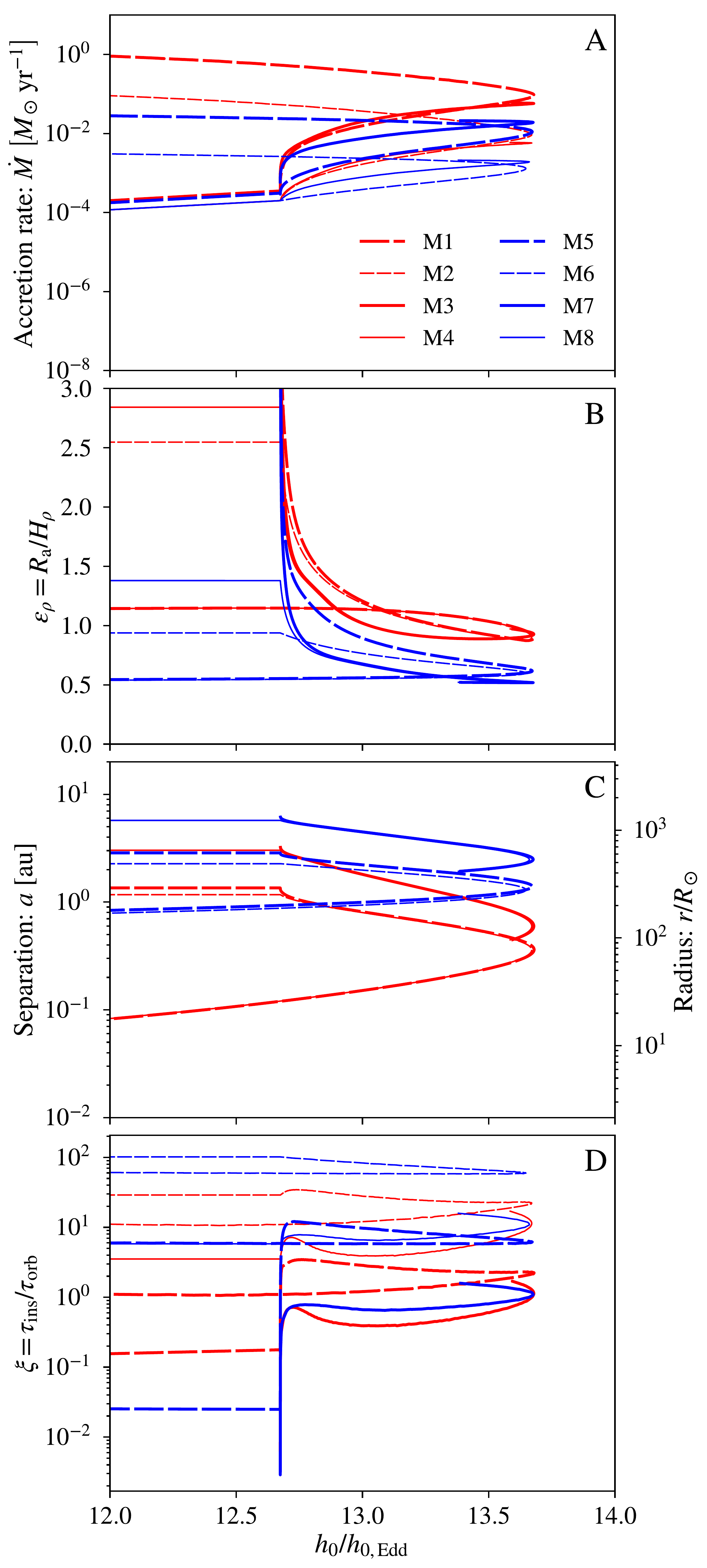}
\caption{\label{fig:vh}
Dependence of model quantities on the strain amplitude.
Panel A: accretion rate.
Panel B: dimensionless density gradient.
Panel C: orbital separation.
Panel D: inspiral parameter.
}
\end{figure}

The orbital separation can be constrained as follows.
We estimate the accretion rate from the GW strain amplitude using Equations~\eqref{eq:strain} and \eqref{eq:quad}. 
From the accretion rate the background density gradient can be computed via Equations \eqref{eq:mdot_norm} -- \eqref{eq:hrh17}.  
The accretion rate as a function of the strain amplitude is plotted in Panel A of Figure \ref{fig:vh}.
During the earlier portions of the inspiral when a majority of the crust remains, higher strain amplitudes imply higher accretion rates.    
While the functional form of the MR15 model admits multiple solutions for $\epsilon_\rho$ given ${\dot M}$, our forward integration with an assumed stellar model selects the correct one. 
Using the data from panels B and H of Figures~\ref{fig:orb1} and \ref{fig:orb2}, we plot in the top panel of Figure \ref{fig:vh} the non-dimensional density gradient, $\epsilon_\rho$, as a function of the strain amplitude relative to the strain amplitude for the NS at its Eddington limit. 
(The NS mass and the distance are used to compute $h_{0,{\rm Edd}}$ here.) 
Time is implicitly a parameter in this plot, and since the strain amplitude increases monotonically with time, both increase from left to right. 
Because of the early steep drop in $\epsilon_\rho$, for a given fixed uncertainty in $h_0/h_{0,{\rm Edd}}$ the best constraints should be obtained later in the evolution, where $\epsilon_\rho$ is smaller.
\par
Assuming a stellar model for the primary, the orbital separation can be constrained from the dimensionless density gradient. 
Figure~\ref{fig:axieps} shows the dependence of $a$ on $\epsilon_\rho$; clearly $a$ is most sensitive to $\epsilon_\rho$ where $\epsilon_\rho$ is better determined from $h_0/h_{0,{\rm Edd}}$. 
Note that while the connection between $h_0/h_{0,{\rm Edd}}$ and $\epsilon_\rho$ depends on the value of $\eta$, the dependence of $a$ on $\epsilon_\rho$ comes from the stellar model only.
We also plot $a$ as a function of the strain amplitude in the middle panel of Figure \ref{fig:vh}, where higher strain amplitudes correspond to lower orbital separations for all of our models.
\par
We list final values of strain amplitude, density gradient, and separation for our models in Table \ref{tab:fac2}. 
Since the stars at the base of the RGB appear to undergo merger, the final separation in these cases is relatively insensitive to the mass of the primary and $\eta$. 
For more evolved stars, the dependence on $\eta$ is essentially undetectable, but changing the mass from 12 to $20M_\odot$ increases the final separation by a factor of 4.3 and decreases the strain amplitude by about 40\%. 
Hence while this technique could be used to measure the amount of CE inspiral, successful application will require some type of constraint on the primary mass.
\begin{table*}
\centering
\caption{\label{tab:fac2} Model quantities at the final timestep.
}
\begin{tabular}{lccccccc}
\hline
     Model  & $h_{0,\rm f}/h_{0,\rm Edd}$
    	    & {\bf $\Omega_{\rm f}/(2\pi) \ ({\rm Hz})$}
            & ${\dot M}_{\rm f} \ (M_\odot \, {\rm yr}^{-1})$
            & $\epsilon_{\rho, \rm f}$
            & $a_{\rm f} \ ({\rm au})$ 
            & ${\dot a}_{\rm f} \ ({\rm au} \ {\rm yr}^{-1})$
            & $\xi_{\rm f}$ \\
\hline
M1 & $7.94$ & $402$ & $2.50$ & $1.29$ & $2.26{\times}10^{-2}$ & $-7.20$ & $2.32$ \\
M2 & $7.76$ & $402$ & $0.261$ & $1.43$ & $2.22{\times}10^{-2}$ & $-0.874$ & $19.5$ \\ 
M3 & $13.6$ & $313$ & $5.91{\times}10^{-2}$ & $0.972$ & $0.444$ & $-2.26$ & $1.71$ \\ 
M4 & $13.6$ & $313$ & $5.91{\times}10^{-2}$ & $0.972$ & $0.444$ & $-0.226$ & $17.1$ \\ 
M5 & $0.00$ & $675$ & $4.74$ & $0.817$ & $1.78{\times}10^{-2}$ & $-4.65$ & $5.22$ \\ 
M6 & $0.00$ & $669$ & $0.485$ & $0.819$ & $1.75{\times}10^{-2}$ & $-0.457$ & $54.1$ \\
M7 & $13.4$ & $324$ & $2.11{\times}10^{-2}$ & $0.521$ & $1.93$ & $-1.59$ & $1.59$ \\ 
M8 & $13.4$ & $324$ & $2.11{\times}10^{-2}$ & $0.521$ & $1.93$ & $-0.159$ & $15.9$ \\ \hline 
\end{tabular}
\end{table*}
\subsection{Constraining the Orbital Decay Rate}
\label{sec:decay}
As we have seen, the inspiral parameter $\xi$ is a measure of how rapid the inspiral is compared to the orbital period. 
A measurement of this parameter would provide information about the CE drag mechanism as well as the envelope structure. 
This is another area in which electromagnetic observations can give us very little information during the CE phase, but combining pre-CE electromagnetic constraints with the GW emission during common envelope can potentially distinguish between plunging and non-plunging scenarios.
\par
\begin{figure}
\centering
\includegraphics[width=\columnwidth]{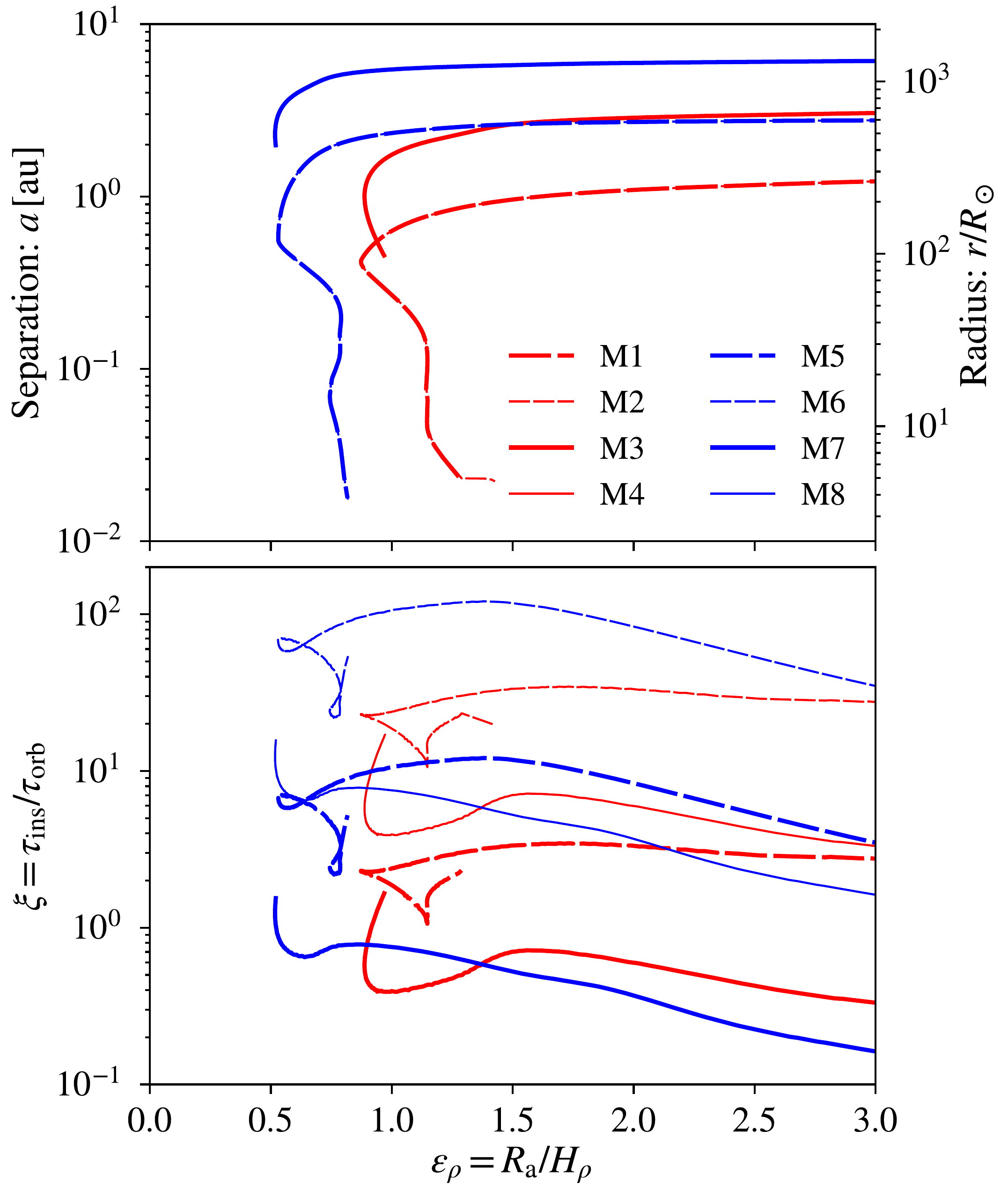}
\caption{\label{fig:axieps}
Top panel: orbital separation vs. $\epsilon_\rho$.  
Bottom panel: inspiral parameter vs. $\epsilon_\rho$.}
\end{figure}
\begin{figure}
\centering
\includegraphics[width=\columnwidth]{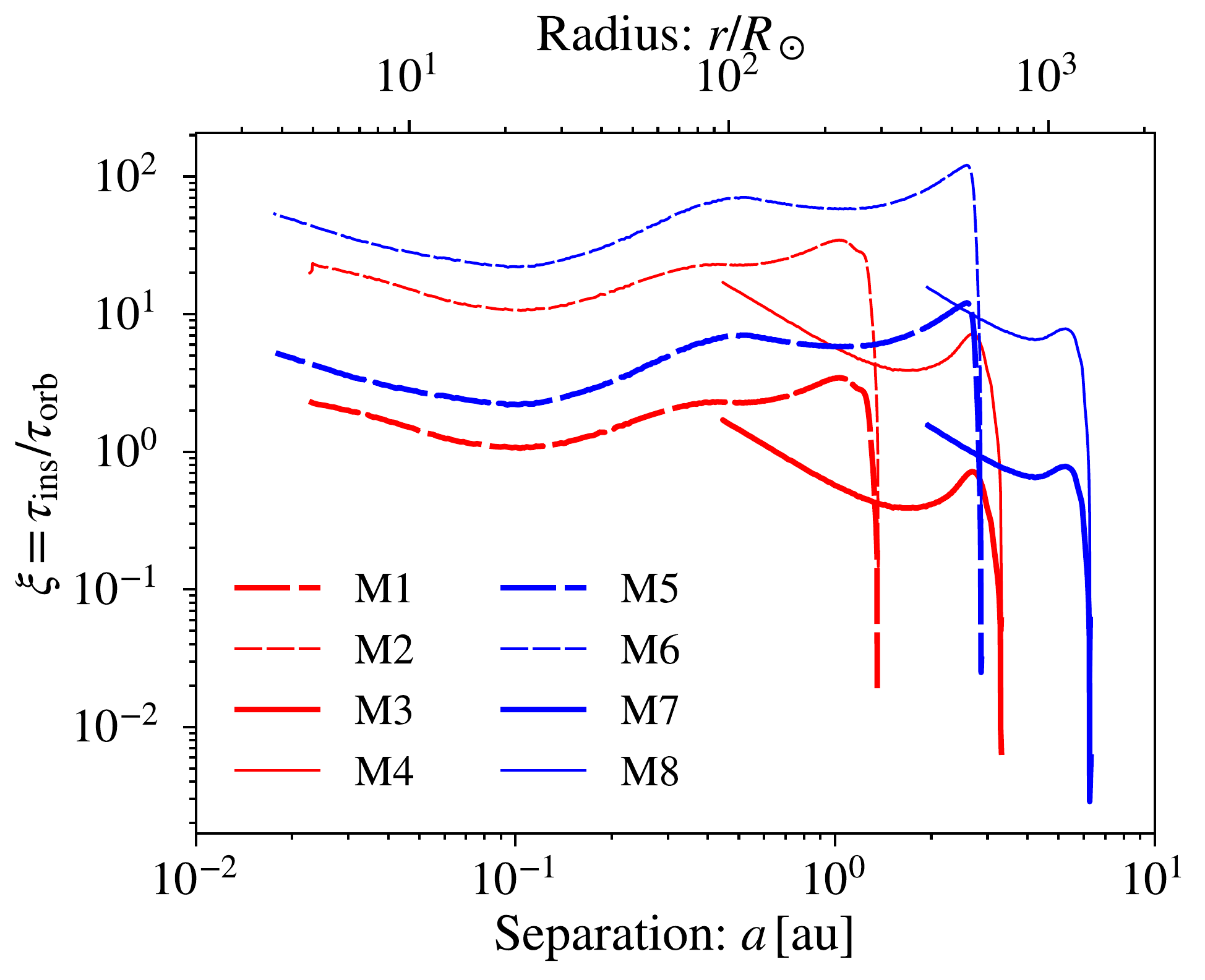}
\caption{\label{fig:xia}
Inspiral parameter, $\xi$, as a function of orbital separation for our models. 
}
\end{figure}

Estimating $\xi$ requires an estimate of $a$ (\S~\ref{sec:orbsep}), to determine the orbital period $\tau_{\rm orb}$, and an estimate of ${\dot a}$, to determine the inspiral timescale $\tau_{\rm ins}$. In both cases, we first obtain $\epsilon_\rho$ from $h_0/h_{\rm 0,Edd}$ as described in the previous section. We then use Equation~\eqref{eq:dragactual} to determine the drag force and Equations \eqref{eq:adot} -- \eqref{eq:vinf} to compute ${\dot a}$. (As with the accretion rate, the functional form of the MR15 drag force model allows multiple solutions for $\epsilon_\rho$, but an assumed stellar model lets us choose the correct one.) 
The inspiral parameter is plotted as a function of the strain amplitude in Panel D of Figure \ref{fig:vh}, as a function of $\epsilon_\rho$ in the bottom panel of Figure \ref{fig:axieps}, and as a function of orbital separation in Figure \ref{fig:xia}. 
We also tabulate values at the end of the inspiral in Table \ref{tab:fac2}.
As expected from the orbital evolution, the final orbital decay rates for the models at the base of the RGB are greater in magnitude than for the more evolved models. 
However, the final inspiral parameter values are smaller at the tip of the RGB, though they are still greater than 1. 
At face value, this seems inconsistent with the apparent merging behavior of the less evolved primaries. 
However, $\xi$ is more rapidly increasing for the more evolved cases. If the integration is carried beyond the termination criterion, $\xi$ continues to increase, while for the stars at the base of the RGB it drops sharply.
\par
As previously mentioned, reducing the accretion rate increases the inspiral timescale, which increases $\xi$, as seen in the $\eta = 0.1$ models relative to the $\eta=1$ models.
Reducing the accretion rate also reduces the strain amplitude, which shifts the model curves to the left in the $\xi-h_0$ plane. Early in the orbital evolution, $\xi\ll 1$ and the quasi-circular approximation is invalid; we have noted that this is due to uncertainties in the outer envelope structure.
During the middle stages of evolution where these uncertainties begin to affect $\epsilon_\rho$ less, the inspiral parameter depends weakly on the strain amplitude. However, at the end where the evolution resolves into stabilization as a close binary or merger, $\xi$ undergoes a sharp uptick. In the non-merger cases this is particularly pronounced, while in the merger cases it is followed by a sharp drop when integrating past the termination criterion. This behavior is nearly independent of $\eta$. This abrupt change in behavior offers a (model-dependent) way to distinguish the two scenarios.

\subsection{Detecting NS-CE systems with gravitational waves} \label{sub:phase}
\begin{figure*} 
\centering
\includegraphics[width=0.9\textwidth]{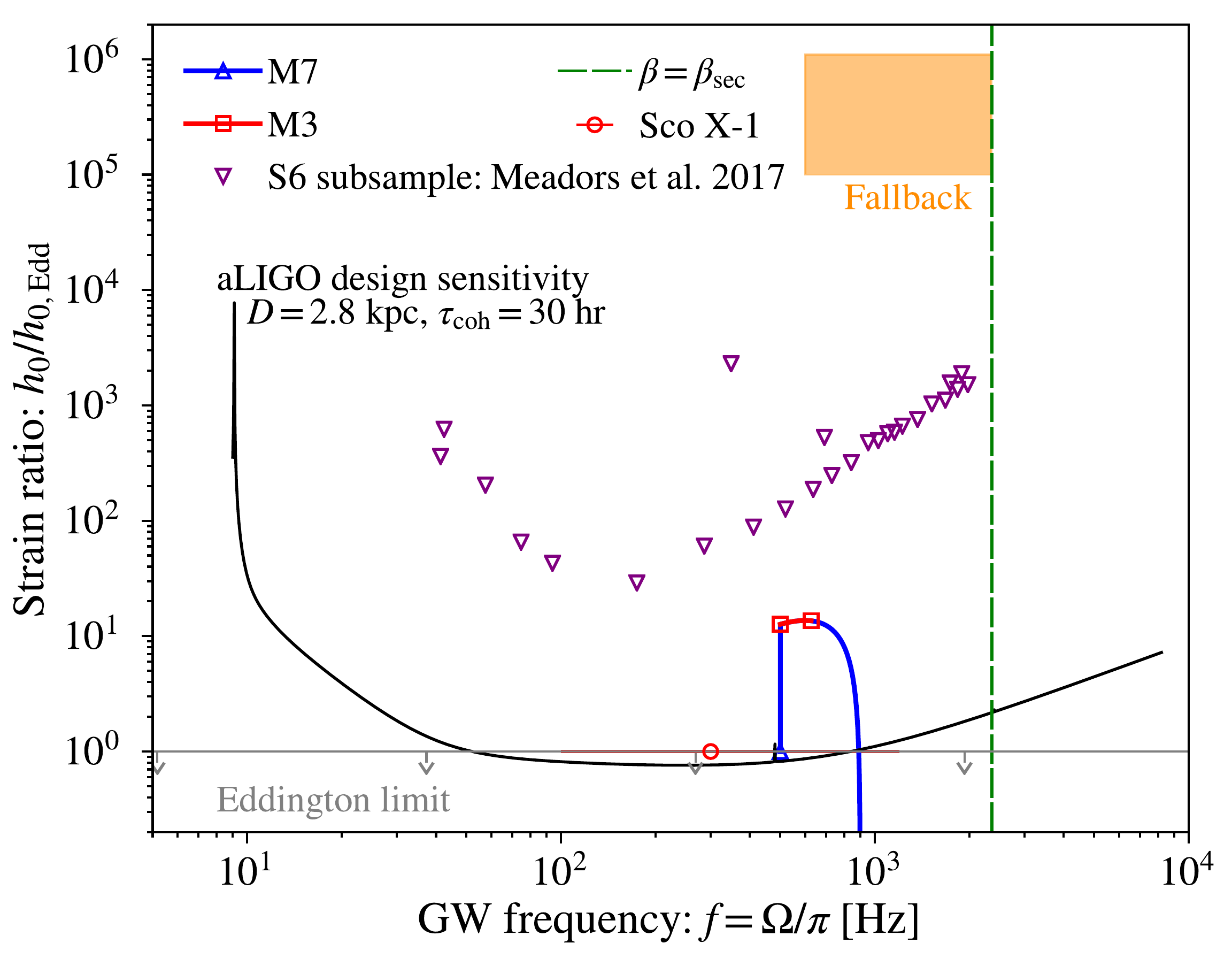}
\caption{\label{fig:phase_space} Strain-amplitude phase space. 
Strain amplitude relative to the strain amplitude for an Eddington-accreting NS as a function of both the accretion rate and GW frequency of the spinning NS at the same distance as Sco X-1, $D = 2.8 \ {\rm kpc}$. 
The Eddington limit is plotted as a gray line with lower arrows.
The lower limit for hyper-critical accretion is plotted as the purple line with upper arrows.
The dashed green line is the threshold of secular instability.
The fallback regime is plotted as the orange region.
The black curve is the aLIGO design sensitivity curve \citep{ligocurve} integrated at a coherence time of $\tau_{\rm coh} = 30 \, {\rm hr}$ for each frequency value, i.e., $N$ continuous-waveform searches results in a total integration time of $N{\times}\tau_{\rm coh}$.
The purple points are a subsample from \cite{meadors_searches_2017} of the 95\%-confidence upper limits to the strain amplitude relative to a torque-balance model using data from the S6 observing run.  
We also plot model ${\rm M}7$ as the blue line with triangle markers and model ${\rm M}3$ as the red line with square markers.
}
\end{figure*}

Any accreting, spinning NS with an asymmetry with respect to its rotation axis is a GW source that may be detectable in the aLIGO band.
In Figure~\ref{fig:phase_space} we compare the GW emission properties of NS-CE systems 
 from models ${\rm M}3$ and ${\rm M}7$ with LMXBs accreting at the Eddington limit, NSs undergoing fallback accretion in supernovae, a subsample of strain-ratio upper limits from \cite{meadors_searches_2017}, and expected strain-ratio upper limits for aLIGO, assuming a source distance of 2.8~kpc. 
As we have seen, even when the density gradient is steep and we make conservative assumptions about the accretion efficiency, NS-CE systems can be about 10 times louder than LMXBs like Sco X-1.
However, fallback accretion rates have been estimated to be ${\dot M}_{\rm fall} \sim 10^{-4} - 10^{-2} M_\odot \ {\rm s}^{-1}$ over a timescale of ${\cal O} (\sim 10^2 \ {\rm s})$
(\cite{macfadyen_supernovae_2001,zhang_fallback_2008}; PT12).
Such accretion rates can exert an accretion torque large enough to spin up the NS up to the point of secular instability, $\beta \sim \beta_{\rm sec}$. 
Thus fallback-accreting systems should be up to {\bf ${\sim}10^{3-4}$} times louder than NS-CE systems. Their characteristic frequencies should reach higher values than for X-ray binaries, but may overlap the binary range before reaching secular instability. However, since the NS-CE systems have a GW modulation timescale ranging from  $\sim 1$ month - 100 yr, it should be easy to distinguish them from fallback accretion events.
\par
For the aLIGO design sensitivity curve, we have taken a coherent integration time of $\tau_{\rm coh} = 30 \, {\rm hr}$, similar to what the Einstein@Home uses for their continuous GW searches (e.g., \cite{ligo_scientific_collaboration_einsteinhome_2009,abbott_results_2016}). 
The coherence time is for an individual frequency such that $N$ continuous-waveform searches within a given interval of the aLIGO band results in a total integration time of $N {\times} \tau_{\rm coh}$. 
If we use its value in the range of frequencies relevant to X-ray binaries, hyper-accreting NS-CE systems should almost be detectable at 2.8 kpc once aLIGO achieves design sensitivity and the loudest merging NS-CE systems may already be detectable at ten times this distance.  
If the NS has achieved the maximum quadrupole moment that it can sustain, the maximum distance at which NS-CE systems would have strain amplitudes comparable to Sco X-1 can be estimated as
\begin{equation}
D \approx 62 \, {\rm kpc} \left(\frac{h_0}{3.5{\times}10^{-26}}\right)^{-1} \left(\frac{P}{1.4 \, {\rm ms}}\right)^{-2} \left(\frac{Q_{\rm  max}}{2{\times}10^{39} \ {\rm g} \, {\rm cm}^2}\right) \ .
\end{equation}
Thus, by the time Sco X-1 becomes detectable via gravitational waves, hyper-accreting NS-CE systems with ${\cal O}(1 \ {\rm ms})$ NS spin periods should be detectable as far away as the Magellanic Clouds.
NS-CE systems with larger NS spin periods may be detectable only within the Galaxy.
For example, a NS-spin period of $\sim 4 \ {\rm ms}$ would only be detectable to $\sim 8 \ {\rm kpc}$, i.e., the Galactic center. 
\section{Conclusions} \label{sec:con} 
In this paper, we have shown that accreting neutron stars undergoing common envelope inspiral may be gravitational wave sources detectable in the aLIGO band as far away as the Magellanic Clouds.
To characterize the evolution of NS-CE systems, we performed orbital integrations using $12 M_\odot$ and $20 M_\odot$ primaries produced with MESA, treating the accretion onto the NS with the MR15 model, using the Maclaurin spheroid model to estimate the spin evolution of the NS, and a quadrupole-moment model to account for melting of the NS crust during the inspiral.  
From the range of possible accretion rates relevant to CE evolution, we find that NS-CE systems may be significantly louder GW sources than LMXBs like Sco X-1, which are currently the target of directed searches for continuous GWs.
We also find that the measured GW strain amplitude of the accreting NS may allow for novel constraints on the orbital separation and inspiral parameter in NS-CE systems when combined with pre-CE electromagnetic observations.   
A sustained mountain producing a time-varying quadrupole moment during spin-up produces a GW signal with a slow chirp-like frequency evolution.
The frequency evolution, however, may be slow enough such that the waveform is continuous over an integration time of ${\cal O}(10 \, {\rm hr})$.
Over the course of the inspiral, the NS crust will melt which reduces the maximum sustainable quadrupole moment.
Once the entire NS crust is melted, there is no GW emission from the electron-capture mechanism.  
\par
Our strain amplitude estimates should be regarded as upper limits. 
However, the threshold for hyper-critical accretion yields a strain amplitude that should be detectable with aLIGO to distances as large as ${\sim}62 \, {\rm kpc}$ at ${\cal O} (1 \, {\rm ms})$ spin periods once Sco X-1 becomes detectable. 
Thus there is potential for extragalactic NS-CE sources to be detected and characterized with aLIGO. In the absence of such detections, we would have to conclude that the accretion may be melting the crust before GW amplitude increases to a detectable level
or the rate of NS-CE events per galaxy is low.
\par
In a future paper we will examine further using 3D simulations some of the assumptions we have adopted here. In particular, plunging phases of CE evolution may produce rapid changes in the GW signal relative to the integration time that may prove useful as diagnostic features. 
The final separation and the condition that defines it are not well-determined yet, but it clearly plays an important role in setting the maximum GW luminosity of NS-CE systems. 
In stars for which the thermal adjustment and dynamical timescales are not too dissimilar, such as the massive stars considered here, the response of the primary to the inspiral cannot be neglected and may determine the final separation and thus maximum GW luminosity. 
Finally, uncertainties in the outer envelope structure clearly matter in the early phases of NS-CE evolution with massive primaries, and further improvements can be sought in this area.
\acknowledgements
We thank the reviewer for providing helpful comments that improved this paper.
AMH is supported by the DOE NNSA Stewardship Science Fellowship.
PMR acknowledges support from the National Science Foundation under grant AST 14-13367. Portions of this work were completed at the Kavli Institute for Theoretical Physics, where it was supported in part by the National Science Foundation under grant NSF PHY-1125915.
We thank Vicky Kalogera, Frank Timmes, Krzysztof Belczynski, Grzegorz Wicktorowicz, Pablo Marchant, Imre Bartos, Daniel Siegel, Konstantin Postnov, Aleksandr Tutukov, Brian Fields, and Lars Bildsten for helpful comments and fruitful discussions.
We also thank Samaya Nissanke for reviewing this work as a LVC reviewer.
The LIGO document number for this work is P1700155. 
\bibliography{srcs,references}
\appendix
\section{Quadrupole moment from electron captures in an isothermal atmosphere} \label{appa}
Following \citep{bildsten_hydrogen_1998}, the mass accretion rate is balanced by the electron capture rate for a given target nuclei in a steady state by
\begin{equation}
{\dot m} \frac{{\rm d}X}{{\rm d}y} = - X R_{\rm ec} \ ,
\end{equation}
where ${\dot m}$ is the mass accretion rate per unit area, $X$ is the concentration of the target nuclei, and $y = \int \rho(z) \, {\rm d}z$.
At the Eddington accretion rate, the mass-accretion rate per unit area is
\begin{equation}
{\dot m}_{\rm Edd} = \frac{\mu_{\rm e}m_{\rm p} c}{\sigma_{\rm T} R_{\rm NS}} = 7.5{\times}10^4 \mu_{\rm e} \left(\frac{R_{\rm NS}}{10 \ {\rm km}}\right) \ {\rm g} \, {\rm cm}^{-2} \, {\rm s}^{-1} \ ,
\end{equation}
where $\mu_{\rm e}$ is the mean molecular weight per electron and $m_{\rm p}$ is the mass of the proton.  
We re-express the above equation in terms of the crust depth
\begin{equation}
\int_{1}^{X} \frac{{\rm d}X'}{X'} = - \int_0^{\Delta z} \frac{R_{\rm ec}}{\dot m} \rho(z) \ {\rm d}z \ .
\end{equation}
The density jump across the electron-capture layer is of order $\Delta \rho/\rho \approx 10\%$ \citep[e.g.,]{ushomirsky_deformations_2000}, so we take an average density: $\int_0^{\Delta z} \rho (z) \, {\rm d}z \approx \bar{\rho}\Delta z$.  
Solving for the thickness of the electron-capture layer, we obtain
\begin{equation} \label{eq:zr}
\Delta z = - \frac{{\dot m}\ln (X)}{\bar{\rho} R_{\rm ec}} \ ,
\end{equation}
The height of mountains generated by electron-captures is of order the electron-capture layer and has an implicit temperature dependence: 
\begin{equation} \label{eq:dz}
\Delta z = \frac{\partial \Delta z}{\partial T} \Delta T \ ,
\end{equation}
where the temperature variation
\begin{equation}
\Delta T = \frac{\partial T}{\partial x} \Delta x \ ,
\end{equation}
may arise from local asymmetric accretion flow, buried magnetic fields, local compositional variation, etc. 
Taking the temperature derivative of Eq.~\eqref{eq:zr}, we obtain
\begin{equation} \label{eq:dT}
\frac{\partial \Delta z}{\partial T} = \frac{{\dot m} \ln (X)}{\rho R_{\rm ec}^2} \frac{\partial R_{\rm ec}}{\partial T} \ .
\end{equation}
From expressions \eqref{eq:dz}-\eqref{eq:dT}, lateral temperature gradients on the NS surface along with the temperature-dependence of the electron-capture rates produces mountains with height ${\cal O}(\Delta z)$. 
As long as these mountains are unaligned with the NS spin axis, the generated quadrupole moment is
\begin{equation}
Q = \varepsilon I = \frac{a-b}{\frac{1}{2}(a+b)} \frac{1}{5} M_{\rm NS} (a^2 + b^2) \ .
\end{equation}
%
where
\begin{subequations}
\begin{align}
a &= R_{\rm NS} + \frac{1}{2}\Delta z \ , \\
b &= R_{\rm NS} - \frac{1}{2}\Delta z \ .
\end{align}
\end{subequations}
The ellipticity is
\begin{equation}
\varepsilon = \frac{\Delta z}{R_{\rm NS}}
\end{equation}
and evaluating the mass quadrupole moment
\begin{equation}
Q = \varepsilon I = \left(\frac{\Delta z}{R_{\rm NS}}\right) \frac{1}{5} M_{\rm NS} \left[\left(R_{\rm NS} + \frac{1}{2}\Delta z\right)^2 + \left(R_{\rm NS} - \frac{1}{2}\Delta z\right)^2\right] \ ,
\end{equation}
and taking $\Delta z \ll R_{\rm NS}$, we ignore terms that are $\ge{\cal O}(\Delta z^2)$ such that $Q$ reduces to
\begin{equation}
Q = \varepsilon I \simeq \frac{2}{5} M_{\rm NS} R_{\rm NS} \Delta z \ .
\end{equation}
%
%
%
We plug in the expression for $\Delta z$ (Eq. \eqref{eq:dz}) to obtain
\begin{equation}
Q = \frac{2}{5} M_{\rm NS} R_{\rm NS} \frac{\partial \Delta z}{\partial T} \Delta T \ .
\end{equation}
We rearrange to get
\begin{equation}
\frac{\partial \Delta z}{\partial T} = \frac{5Q}{2 M_{\rm NS}R_{\rm NS}\Delta T} \ .
\end{equation}
We then plug in Eq.~\eqref{eq:dT} to obtain the required temperature dependence of the electron-capture rate for a desired quadrupole moment $Q_0$
\begin{equation} \label{eq:cond}
\frac{\partial R_{\rm ec}}{\partial T} \ge \frac{\rho R_{\rm ec}^2}{{\dot m}\ln(X)} \frac{5 Q_0}{2 M R_{\rm NS}\Delta T} \ .
\end{equation}
If the left-hand side of the above expression is less than the right-hand side, then the generated quadrupole moment is less than the desired one: $Q < Q_0$.   
For the hypercritical regime, we consider a target nuclei of $^{56}{\rm Ni}$ with $R_{\rm ec}$ and $\partial R_{\rm ec}/\partial T$ obtained from \cite{nabi_gamow_2005}, a remaining concentration $X=0.1$, ${\dot m} = 10^4{\dot m}_{\rm Edd}$, $Q_0 = 4{\times}10^{39} \ {\rm g} \, {\rm cm}^2$ , $T = 10^{10} \ {\rm K}$, ${\bar \rho} = 10^9 \ {\rm g} \, {\rm cm}^{-3}$, and $\Delta T = 0.01 T$, we find that Eq.~\eqref{eq:cond} is satisfied, which suggests that electron captures may be a relevant mechanism for generating GW emission for NS-CE events.  
Evaluating Eq. \eqref{eq:cond}, for the Eddington case with hydrogen and using $R_{\rm ec}$ and $\partial R_{\rm ec}/\partial T$ from \cite{bildsten_hydrogen_1998}, ${\dot m} = {\dot m}_{\rm Edd}$, $T = 10^{9} \ {\rm K}$, ${\bar \rho} = 10^7 \ {\rm g} \, {\rm cm}^{-3}$, and other values the same as the previous case, Eq. \eqref{eq:cond} is satisfied.  
%
%
%
%
%
\end{document}